\documentclass[letterpaper,twocolumn,10pt]{article}
\usepackage{usenix-2020-09}
\usepackage{tikz}
\usepackage{graphicx}
\usepackage{float}
\usepackage{subfig}
\usepackage{colortbl}
\usepackage{color}
\usepackage{amsmath}
\usepackage [ruled,linesnumbered,algo2e] {algorithm2e}
\usepackage{algorithm}
\usepackage{algpseudocode}
\usepackage{makecell}
\usepackage{cleveref}
\usepackage{enumitem}
\usepackage[available,functional]{usenixbadges}

\setitemize{noitemsep,topsep=0pt,parsep=0pt,partopsep=0pt,leftmargin=*}
\usepackage{filecontents}
\usepackage{amsmath,amssymb,amsfonts}
\usepackage{textcomp}
\usepackage{xcolor}
\usepackage[]{url}

\begin{document}

\date{}

\title{\Large \bf \SystemName: Automatically Pushing the Envelope of Multi-GPU System\\ for Billion-Scale GNN Training}
\author{
 {\rm \hspace{0.3cm}  Jie Sun$^{1}$,  \hspace{0.3cm}  Li Su$^{2}$,  \hspace{0.3cm}  Zuocheng Shi$^{1}$, \hspace{0.3cm}  Wenting Shen$^{2}$, \hspace{0.3cm}  Zeke Wang$^{1}$} \\ {\rm \hspace{0.3cm} Lei Wang$^{2}$, \hspace{0.3cm}  Jie Zhang$^{1}$, \hspace{0.3cm}  Yong Li$^{2}$, \hspace{0.3cm}  Wenyuan Yu$^{2}$, \hspace{0.3cm}  Jingren Zhou$^{2}$, \hspace{0.3cm}  Fei Wu$^{1,3}$}\\
 \small $^{1}$ Collaborative Innovation Center of Artificial Intelligence, Zhejiang University, China \\
 \small $^{2}$ Alibaba Group\\
 \small $^{3}$ Shanghai Institute for Advanced Study of Zhejiang University, China
\\
} 

\newcommand{\SystemName}{{Legion}\xspace}
\newcommand{\sunjie}[1]{\textcolor{blue}{#1}}

\maketitle{}

\pagestyle{empty}
\begin{abstract}

Graph neural network(GNN) has been widely applied in real-world applications, such as product recommendation in e-commerce platforms and risk control in financial management systems. Several cache-based GNN systems have been built to accelerate GNN training in a single machine with multiple GPUs. However, these systems fail to train billion-scale graphs efficiently, which is a common challenge in the industry. In this work, we propose \SystemName{}, a system that automatically pushes the envelope of multi-GPU systems for accelerating billion-scale GNN training. First, we design a hierarchical graph partitioning mechanism that significantly improves the multi-GPU cache performance. Second, we build a unified multi-GPU cache that helps to minimize the PCIe traffic incurred by caching both graph topology and features with the highest hotness. Third, we develop an automatic cache management mechanism that adapts the multi-GPU cache plan according to the hardware specifications and various graphs to maximize the overall training throughput. Evaluations on various GNN models and multiple datasets show that \SystemName{} supports training billion-scale GNNs in a single machine and significantly outperforms the state-of-the-art cache-based systems on small graphs.
\end{abstract}


\section{Introduction}
Graph neural networks (GNNs), such as~\cite{kipf2016semi, hamilton2017inductive,  velivckovic2017graph, chiang2019cluster, zeng2019graphsaint, chen2018fastgcn}, are a class of deep learning algorithms that learn the low-dimensional embedding using the structure and attribute information of graphs. The learned embedding can be further used in machine-learning tasks including node classification and link prediction. GNNs have been successfully applied in many real-world applications, such as recommendation systems in e-commerce platforms, fraud detection and risk control in financial management, and molecular property prediction in drug development~\cite{ying2018graph, liu2021pick, yuapplication, su2022gnn, gilmer2017neural}. Systems such as DGL~\cite{wang2019deep}, PyG~\cite{paszke2019pytorch}, and Graph-Learn~\cite{zhu2019aligraph} are proposed to ease the development and training of GNN models.

It is common to apply GNNs over large-scale graphs in industrial scenarios. For example, in Alibaba's Taobao recommendation system, the user behavior graph contains more than one billion vertices and tens of billions of edges~\cite{zhu2019aligraph}. In addition, as graphs are often skewed, it is infeasible to aggregate all neighboring vertices when training a specific vertex. Sampling-based mini-batch training, such as GraphSAGE~\cite{hamilton2017inductive}, is proposed to extend GNN training to very large graphs. 
In the sampling-based GNN training, there are two key steps of data preparations before training a batch: (1) sampling the multi-hop sub-graph for each vertex in the batch, and (2) extracting the features of vertices in sampled sub-graphs. Systems such as DGL~\cite{wang2019deep} and PyG~\cite{paszke2019pytorch} store the graph data in the CPU memory, prepare the training data of mini-batches using CPUs, and utilize GPUs for model training. As this approach requires transferring the sampled sub-graphs and high-dimension feature data to the GPU for every batch, the end-to-end training throughput is severely limited by the CPU-GPU data transferring bandwidth~\cite{lin2020pagraph, yang2022gnnlab}. In addition, the throughput of graph sampling using CPU is often insufficient to keep up with the throughput of GPU training, especially in multi-GPU machines.

Several cache-based approaches have been proposed to speed up GNN training~\cite{lin2020pagraph, torchquiver, yang2022gnnlab, min2022graph}. As it is the feature data that accounts for a majority of the CPU-GPU data transferring, caching the features of frequently accessed vertices in GPU can significantly reduce the amount of transferred data. To improve the throughput of graph sampling, GPU-based sampling has also been adopted in GNN systems~\cite{wang2019deep, yang2022gnnlab, torchquiver}.

We identify that existing approaches face severe limitations or performance issues in multi-GPU training, particularly when the graph is large.
First, the multi-GPU cache scalability of existing cache-based systems is poor. Some cache-based GNN systems~\cite{yang2022gnnlab, torchquiver} shuffle the training set across all GPUs and replicate an identical feature cache across all GPUs or NVLink cliques\footnote{NVLink clique denotes a group of GPUs where each pair of GPUs are connected with NVLink.} to facilitate data parallel training. The cache capacity is constrained by the memory of a single GPU or NVLink clique (an NVLink clique only consists of two GPUs in some multi-GPU architectures), resulting in poor cache performance when scaling up the number of GPUs (see the experiment in Figure~\ref{fig_comparison_of_four_types}). PaGraph~\cite{lin2020pagraph} partitions the graph using a self-reliant algorithm and caches nodes with the highest in-degree for different partitions in different GPUs, trying to make use of data locality inside each partition. As partitions in PaGraph include the complete L-hop neighbors of their training vertices, there is a significant overlap between the caches of different partitions, resulting in the same duplication issue as the aforementioned cache-based GNN systems.
Second, when adopting GPU-based graph sampling, existing systems manage the graph topology in a very coarse-grained manner: the topology has to be completely stored in a single GPU~\cite{yang2022gnnlab, wang2019deep, torchquiver} or in the CPU memory~\cite{wang2019deep, torchquiver}. The former approach puts a hard limit on the graph scale, and further squeezes the cache capacity for features. The latter storing the topology in the CPU and accessing it from GPU would result in very low utilization of the PCIe bandwidth, as the data access of graph sampling is usually random and fine-grained.

This paper presents \SystemName{}, a GNN system that fully explores the hardware capabilities of modern multi-GPU servers for training large-scale graphs in a single machine. \SystemName{} proposes two key designs to fully exploit the memory space of multi-GPUs for feature and topology cache. First, to avoid cache replication, we propose \textbf{NVLink-aware  hierarchical graph partitioning} technique that helps scale the cache on multi-GPU memory efficiently according to the specific hardware structure. \SystemName{} first partitions the graph with minimal edge-cut and assigns each partition exclusively to an NVLink clique, and then uses hash partition to further map the training vertices to GPUs inside each NVLink clique. Second, we propose a \textbf{hotness-aware unified cache} that manages both the feature and topology cache in a vertex-centric data structure. We enable an NVLink-enhanced cache space for the unified cache and prioritize the topology and features with the highest hotness to fill the cache, so as to improve the multi-GPU memory utilization. 

The above designs pose a new challenge to \SystemName{}. 
Given a fixed size of GPU memory, it is hard to manually decide the optimal fractions of topology and feature cache such that the overall training throughput is maximized. To solve the challenge, we propose an \textbf{automatic cache management} mechanism. Specifically, we build a cost model in the mechanism to evaluate the key factor to the overall throughput, i.e., PCIe traffic, of both graph sampling and feature extraction in the training phase, which is used to guide the allocations of cache spaces for graph topology and feature. Overall, the three key designs in \SystemName{} enable automatic caching optimization and full utilization of hardware capability of various modern GPU servers. 
Experiments show that \SystemName{} can outperform state-of-the-art cache-based GNN systems up to $4.32\times$. 

In summary, the contributions of this paper include:

\begin{enumerate}[nosep,leftmargin=*]
\item We propose an NVLink-aware hierarchical graph partitioning technique that helps minimize cache replication between NVLink cliques and extends the threshold of cache capacity beyond the limit of an NVLink clique.
\item We propose a hotness-aware unified cache to store topology and features with the highest hotness in GPU memory, so as to improve the GPU memory utilization. 
\item We present an automatic cache management mechanism that searches for the optimal cache plan without requiring extra knowledge of hardware specifications and GNN performance details from users.
\item We implement \SystemName{} that fully explores the hardware capabilities of multi-GPU systems targeting billion-scale GNN training, not supported by existing cache-based GNN systems, in a single server. 
\end{enumerate}

\vspace{2ex}

\vspace{-2ex}
\section{Preliminaries}
\label{preliminiarytxt}

In this section, we introduce the basic concept of GNN and the workflow of mini-batch GNN training.

\subsection{Graph Neural Networks}
Given a graph  $ G = (V, E) $, where each vertex is associated with a vector of data as its features $ X_v, v \in V $, Graph Neural Networks(GNNs) learn a low-dimensional embedding for each target vertex by stacking multiple GNNs layers $L$. For each layer $l, l \in L$, vertex $v$ updates its activation by aggregating features or hidden activations of its neighbors $ N(v), v \in V  $:
\vspace{-4ex}





\begin{equation} \label{agg_update}
\begin{split}
    a_v^l &= AGGREGATE^l({h_u^{l-1}|u \in N(v)}) \\
    h_v^l &= UPDATE^l(a_v^l, h_v^{l-1})
\end{split}
\end{equation}

\subsection{Mini-batch GNNs Training}

\begin{figure}[t]
        \centering
        \includegraphics[width=3.35in]{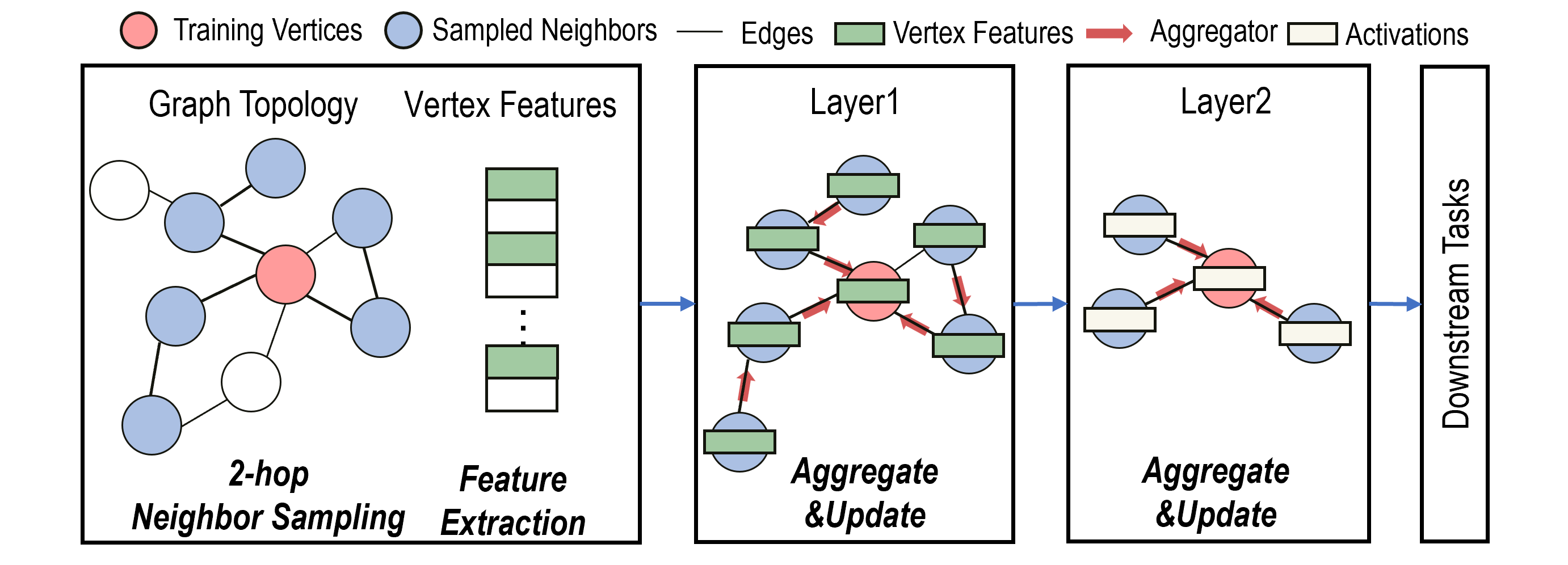} 
	\vspace{-3.5ex}	
	\caption{The workflow of 2-hop GraphSAGE training. } 
	\vspace{-2ex}
	\label{graphsage_training} 
\end{figure} 

Mini-batch training is a practical solution for scaling GNN training on very large graphs. Neighbor sampling is used to generate mini-batches, allowing sampling-based GNN models to handle unseen vertices. For example, GraphSAGE\cite{hamilton2017inductive} samples multiple hops of neighbors for training as shown in Figure~\ref{graphsage_training}.
The workflow of GraphSAGE training follows a vertex-centric computation paradigm including the following steps: 1, selecting a mini-batch of training vertices from the training set. 2, uniformly sampling the multiple hops of fixed-size neighbors for each training vertex. 3, extracting the features of the sub-graph consisting of the training vertices and their neighbors to generate the mini-batch training data. Finally, performing \textit{AGGREGATE} and \textit{UPDATE} according to Equations~\ref{agg_update}, as well as executing the forward and backward propagation to update the model parameters.

\vspace{-2ex}

\section{Observation and Motivation}
When training large-scale graphs whose size exceeds the capacity of GPU memory on a multi-GPU server, the major performance bottleneck becomes the data movement from CPU to GPUs under the constraint of PCIe bandwidth. To this end, existing works~\cite{wang2019deep, yang2022gnnlab, torchquiver} intend to relieve the PCIe bandwidth bottleneck by caching the hottest graph features on GPU memory. Though these cache-based approaches significantly reduce PCIe traffic, we still identify two issues of these existing cache-based GNN systems when training large-scale graphs: 1) poor multi-GPU cache scalability, and 2) coarse-grained GPU memory management for graph topology. In the following, we discuss each issue and the corresponding observation that motivates the design of \SystemName. 

\subsection{Multi-GPU Cache Scalability}
\label{clusteringmotivation}

\begin{figure}[t]
    \centering
    \subfloat[2 GPUs per NVLink clique]
    {\includegraphics[width=1.6in]{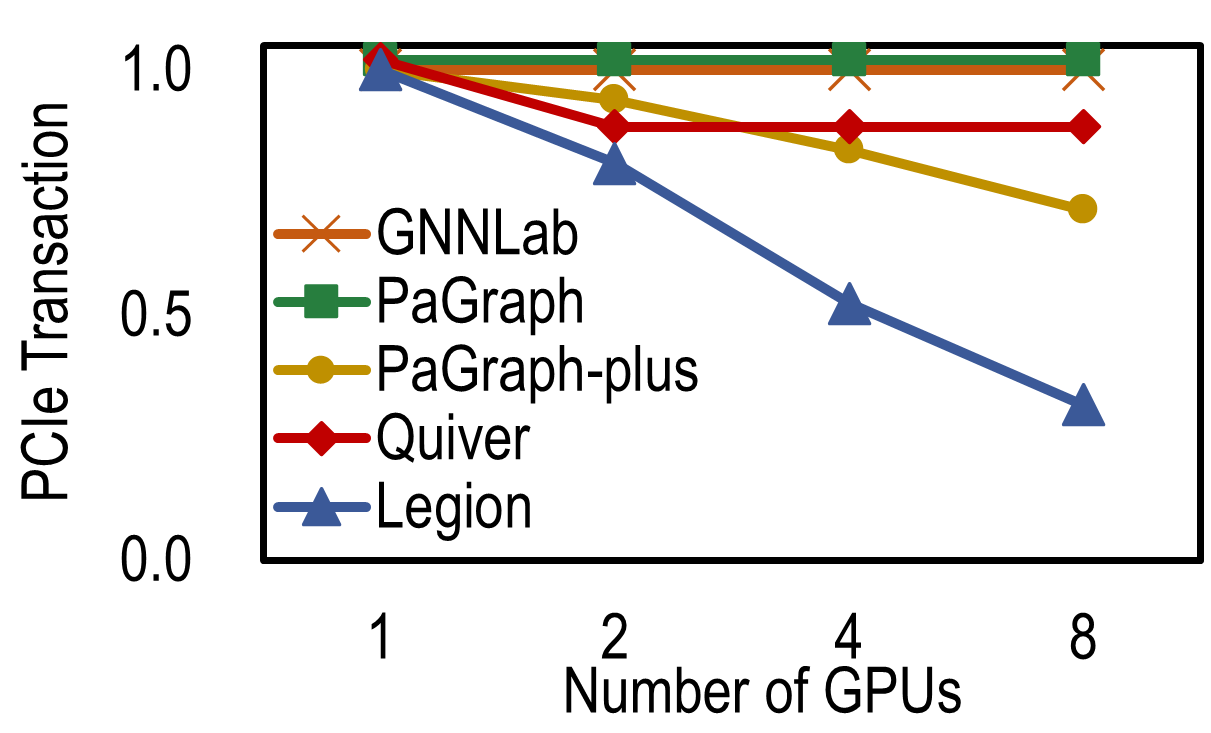} \label{cachescalabilitysiton}}
    \subfloat[4 GPUs per NVLink clique]
    {\includegraphics[width=1.6in]{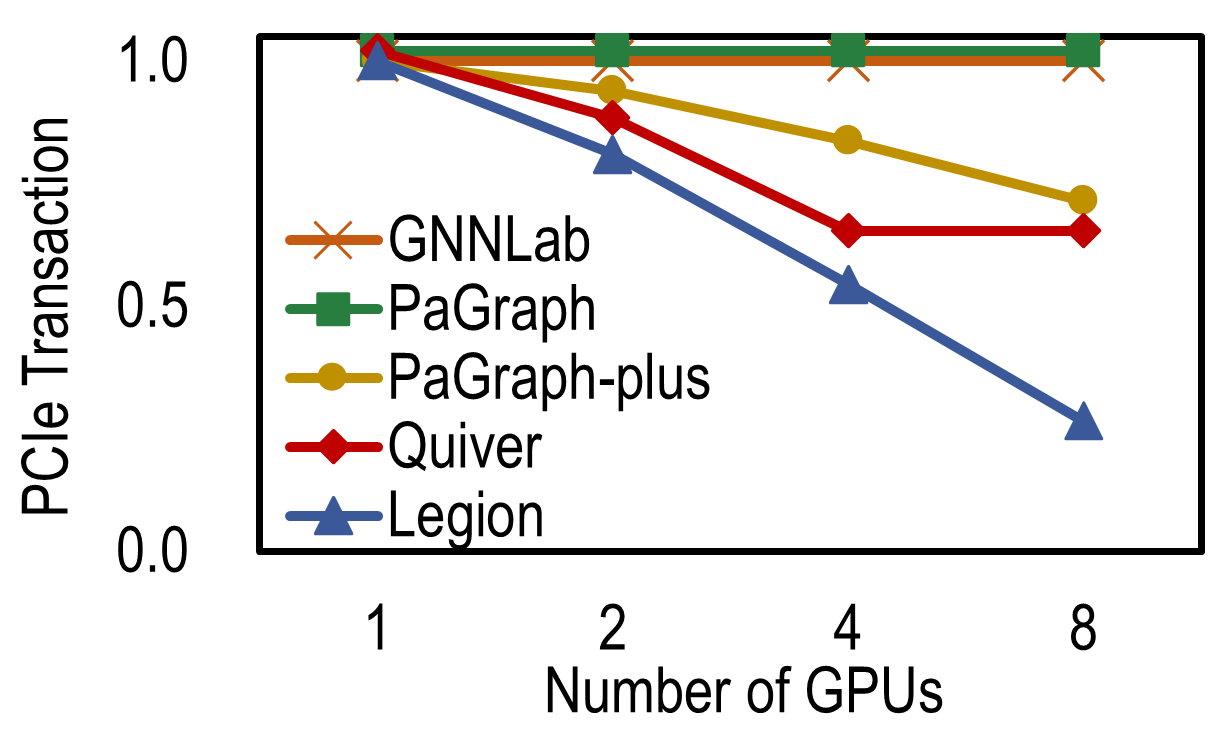} 
        \label{cachescalabilitydgx}} 
	\caption{Comparing the cache scalability of cache-based GNN systems using the Products~\cite{hu2020open} dataset and 2-hop GraphSAGE~\cite{hamilton2017inductive} model in terms of normalized CPU-GPU PCIe transactions. The cache ratio is set to 5\% $|V|$ on every GPU. The tested platforms are Siton (a) and DGX-V100 (b) servers, as shown in Table~\ref{serverstat}.}
 	\vspace{-3ex}
\label{fig_comparison_of_four_types} 
\end{figure}

As feature extraction occupies most of the data transferring from CPU to GPU, cache-based systems like GNNLab~\cite{yang2022gnnlab} maintain a global feature cache for vertices which are more frequently accessed via a pre-sampling phase. As training vertices are globally shuffled among all training GPUs, GNNLab replicates this cache across all GPUs involved in model training. Since a single GPU's memory space is quite limited, the fraction of cached features would inevitably become lower when the graph size grows, resulting in a lower cache hit ratio even on multi-GPU servers. To increase the cache capacity, the cache mechanism in Quiver~\cite{torchquiver}
leverages high-speed NVLinks to support inter-GPU cache between NVLink-connected GPUs. Different from GNNLab, Quiver replicates feature cache between NVLink cliques and averagely hashes the features among GPUs in the same NVLink clique. However, this mechanism could still lead to poor cache scalability, especially when the NVLink clique is relatively small. E.g., the Siton server used in Table~\ref{serverstat} has $4$ NVLink cliques, each of which contains only $2$ GPUs. Figure~\ref{fig_comparison_of_four_types} illustrates that, in systems like Quiver, the PCIe transactions incurred by CPU-GPU data transferring stop decreasing when the number of GPUs is larger than the size of NVLink clique. This result shows that the cache performance in the above GNN systems cannot scale well with the increasing number of GPUs in modern servers.

\begin{figure}[t]
        \centering	\includegraphics[width=2in]{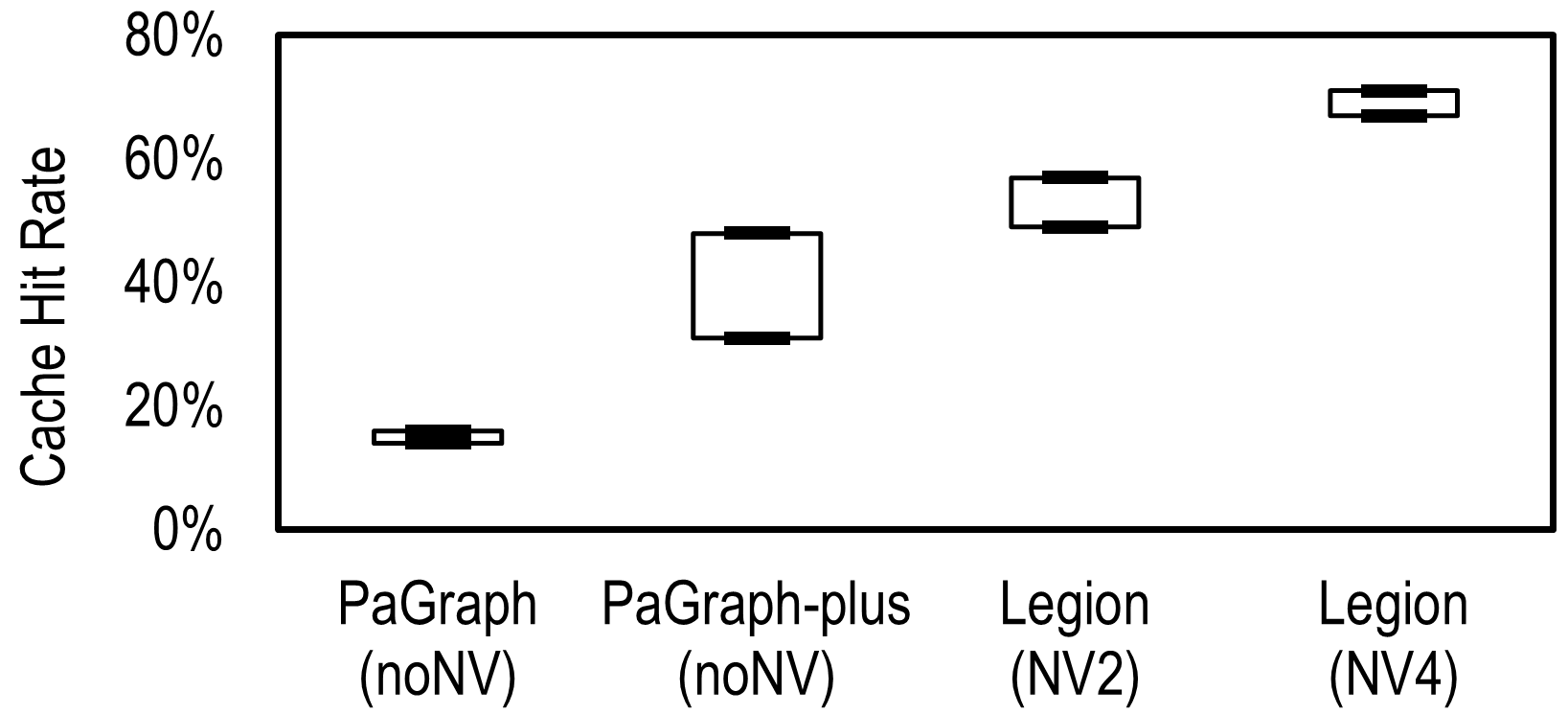} 
        \vspace{-2ex}	
\caption{Cache hit rates of different systems in a server with 8 GPUs. The cache ratio is set to 5\% $|V|$ on every GPU. The graph sampling follows the 2-hop GraphSAGE~\cite{hamilton2017inductive} model's setting using the Products~\cite{hu2020open} dataset. ``NVx" means utilizing NVLink clique with x GPUs.
} 
	\vspace{-1ex}
	\label{loadbalance} 
        \vspace{-1ex}
\end{figure}

To solve the scalability issue incurred by cache replication, PaGraph~\cite{lin2020pagraph} partitions the graph in a self-reliance approach and maintains an independent cache for each partition using an in-degree-based metric on different GPUs. To train an L-layer GNN model, PaGraph extends every partition with redundant vertices and edges to include all the L-hop neighbor vertices for each train vertex in this partition. Each GPU only trains its own partition and synchronizes its local gradients periodically to update the model.
However, the inclusion of the L-hop neighbor vertices leads to heavily duplicated cache contents on all GPUs. Figure~\ref{fig_comparison_of_four_types} shows that the PaGraph exhibits a similar cache performance as GNNLab which adopts the cache replication mechanism. We further implement a PaGraph-plus design to alleviate the cache duplication issue in PaGraph. Specifically, we replace the graph partitioning algorithm in PaGraph with the XtraPulp~\cite{slota2017partitioning} algorithm that minimizes edge-cuts between partitions and adopts a pre-sampling-based hotness metric to select the vertex features to be cached. Although PaGraph-plus achieves higher cache hit rates compared to PaGraph, the
cache hit rates on different GPUs are very unbalanced as different partitions have various graph distributions. 
Figure~\ref{loadbalance} illustrates the load imbalance issue of PaGraph-plus by measuring the cache hit rates of eight GPUs. We observe that the hit rate varies by up to 17\%.

To sum up, for systems that globally shuffle the training vertices among GPUs in every iteration, such as GNNLab and Quiver, cache replication cannot be completely eliminated as each GPU may randomly access any vertex in the entire graph. Whereas the high-speed NVLinks between GPUs can be used to reduce the replication factor and expand the cache capacity. For systems that locally shuffle training vertices in each partition to produce mini-batches for different GPUs, such as PaGraph, the cache replication problem could be alleviated only when the model layer is small (e.g., less than $2$). PaGraph-plus can further reduce cache duplication but faces another issue of unbalanced cache hit rates among GPUs.

\noindent{\textbf{Observation O1: Graph partitioning can be suitably guided by hardware structure. } 
Different from Quiver, GNNLab, PaGraph, and PaGraph-plus do not take advantage of the NVLink between GPUs, which is a common capability in modern multi-GPU servers. As GPUs inside the same NVLink clique can access each other's memory via the low-latency high-throughput NVLink, an NVLink clique can hold the entire cache of a partition, which can be randomly sliced and averagely allocated among GPUs inside a clique. This hardware-coherent design can balance the cache hit ratios between intra-clique GPUs. As the number of partitions is reduced to the number of NVLink cliques, it is more likely that the partitions follow a similar distribution (see the cache hit rate distribution of \SystemName in Figure~\ref{loadbalance}). Inspired by \textbf{O1}, we propose an NVLink-aware hierarchical partitioning to preserve multi-GPU cache scalability in \SystemName(Section~\ref{hiclustering}).
}

\begin{figure}[t]
    \subfloat[PCIe throughput vs. payload size]
    {\includegraphics[width=1.6in]{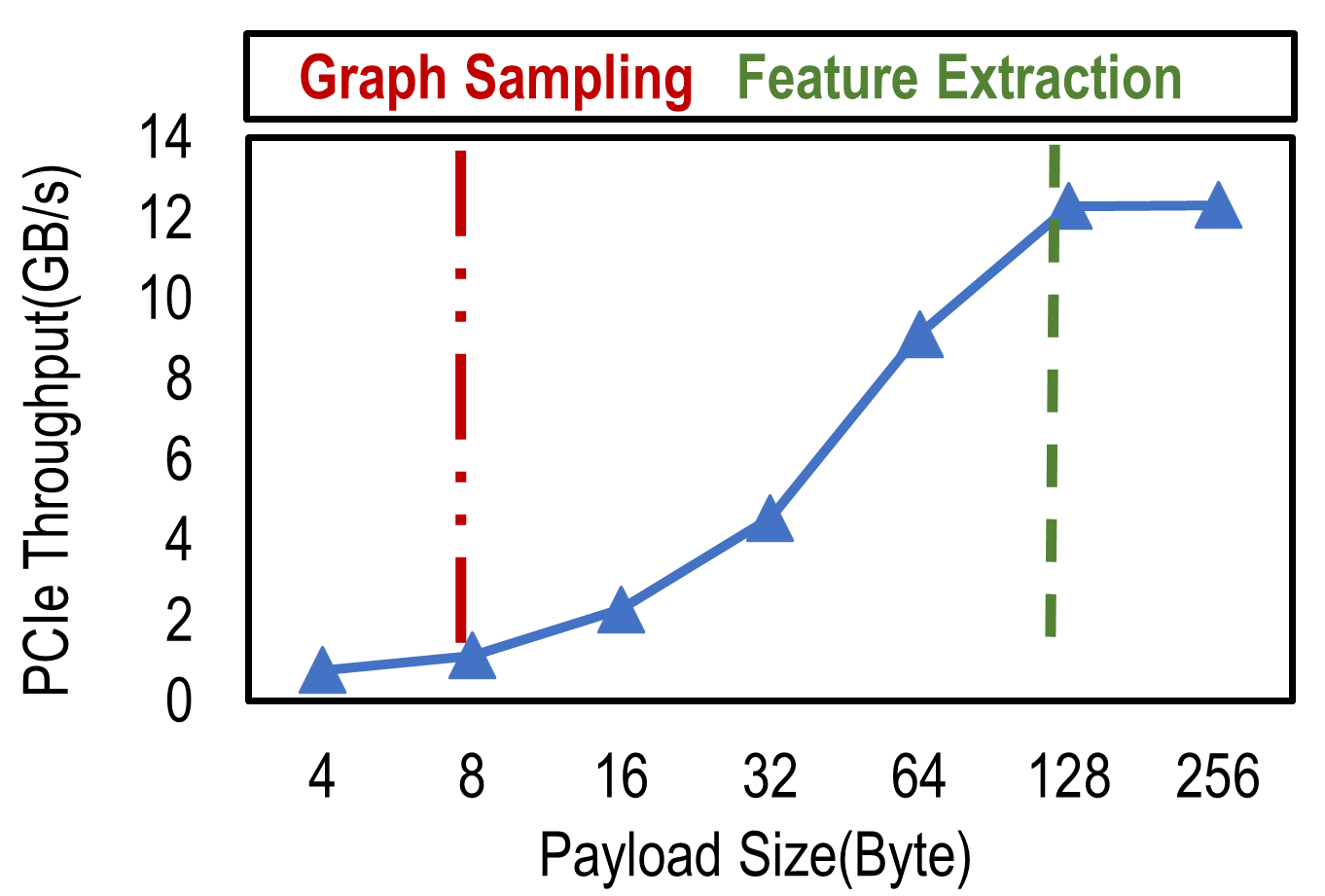} \label{pciethroughput}}
    \subfloat[]
    {\includegraphics[width=1.6in]{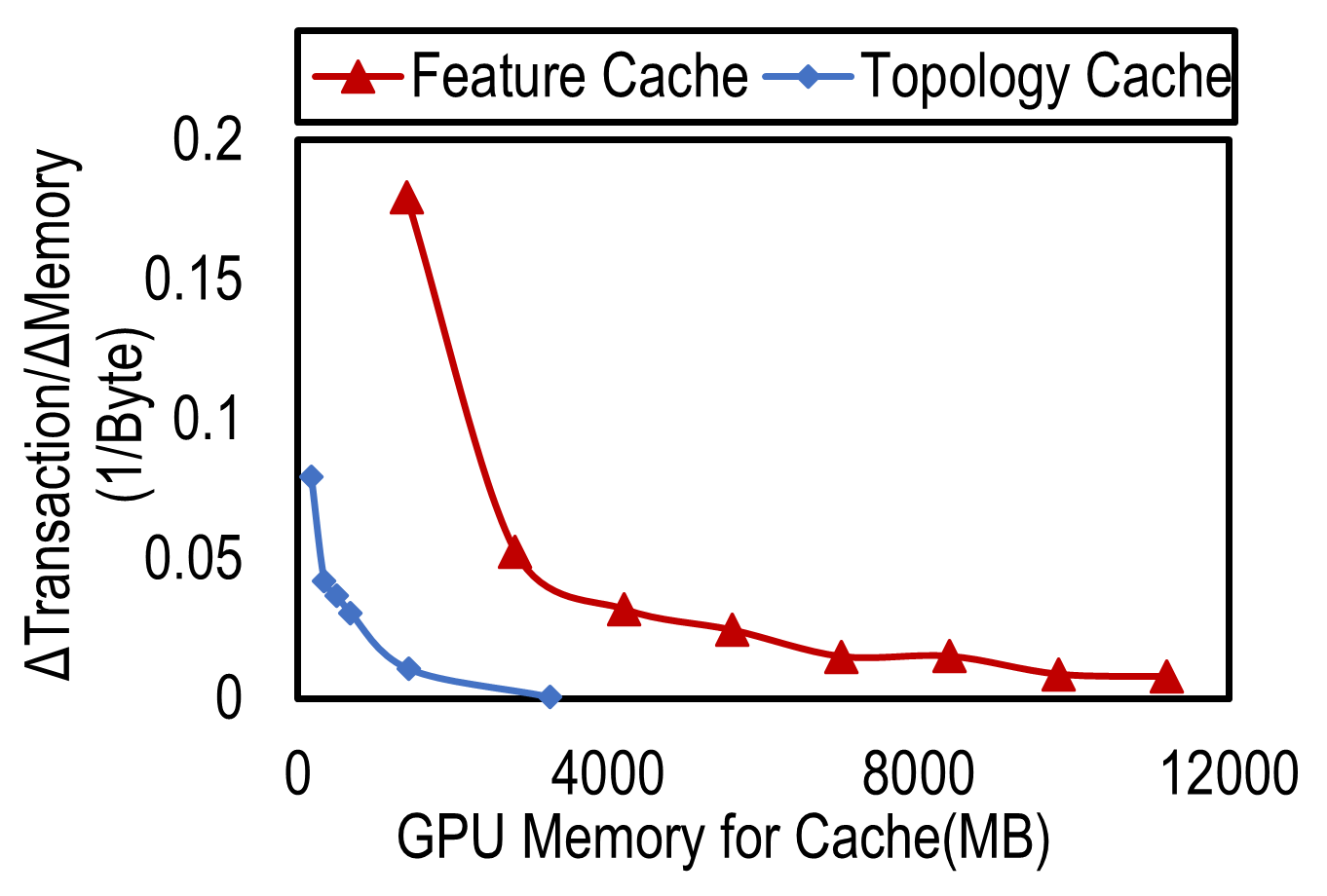} 
        \label{cachewithsize}} 
	\caption{(a) The PCIe 3.0 throughput under different payload sizes of PCIe requests. (b) The PCIe traffic reduction rate for Paper100M with the growth of the cache capacity. The cache is on a single GPU and selected after pre-sampling.}
	\label{motivationexp} 
        \vspace{-2ex}
\end{figure}

\subsection{Coarse-grained GPU Memory Management for Graph Topology}
\label{topologycachemotivation}

In multi-GPU servers, the throughput of CPU-based graph sampling may not catch up with the throughput of GPU-based training. To improve the end-to-end training throughput, recent GNN systems~\cite{wang2019deep, yang2022gnnlab, torchquiver} adopt GPUs to accelerate graph sampling. We observe that all these systems apply a very coarse-grained memory management mechanism for graph topology. In particular, they store the entire graph topology either in CPU memory or in a single GPU, depending on the size of graph topology:
the graph topology is stored in CPU memory when it is too large or exceeds the capacity of a single GPU.
The approach of storing the entire graph topology in a single GPU sets a hard limit on the scale of the graph. For example, a V100 GPU with 16GB memory can store at most $4$ billion edges~\cite{hamilton2017inductive} without considering any other memory usage of feature cache and model training. When storing the graph topology in CPU memory, GPUs can directly access the graph topology via a unified virtual memory address (UVA~\cite{UVAref}) technique. While the data access pattern of graph sampling is usually random and fine-grained. E.g., Figure~\ref{pciethroughput} shows that the PCIe throughput of graph sampling is much lower than feature extraction. A large number of sampling PCIe transactions with small payload sizes will increase the CPU-GPU PCIe contention and lead to low bandwidth utilization.

\noindent{\textbf{Observation O2: The access of graph topology is skewed as graph features.}
Existing cache-based GNN systems~\cite{yang2022gnnlab, lin2020pagraph, torchquiver} only maintain feature cache in GPU to reduce the CPU-GPU communication costs. However, we observe that the performance gain of the per-unit feature cache decreases once the cache capacity exceeds a threshold (see Figure~\ref{cachewithsize}).  We observe that the access of graph topology during graph sampling is also skewed as the access of features. Instead of allocating all the available GPU memory (except for the reservation for model training) for feature cache, it is reasonable to cache a subset of graph topology, i.e., edges of vertices that are frequently accessed during sampling, in the GPU memory to accelerate GPU sampling. Figure~\ref{cachewithsize} shows that a relatively small topology cache can obviously reduce the number of PCIe transactions incurred by GPU sampling. Motivated by \textbf{O2}, we propose a hotness-aware unified cache in \SystemName. Specifically, \SystemName caches both graph topology and graph features with the goal of minimizing CPU-GPU communication overhead (see Section~\ref{unifiedcachetxt}). Under the capacity limit of  GPU memory, it is difficult to manually decide the optimal fractions of topology and feature cache. \SystemName solves this challenge with an automatic cache management mechanism, which can generate the optimal cache plan without requiring knowledge of hardware specifications from users.

\vspace{1ex}

}

\vspace{-2ex}

\section{Design of \SystemName{}}
\begin{figure*}[t]
        \centering
	\includegraphics[width=7in]{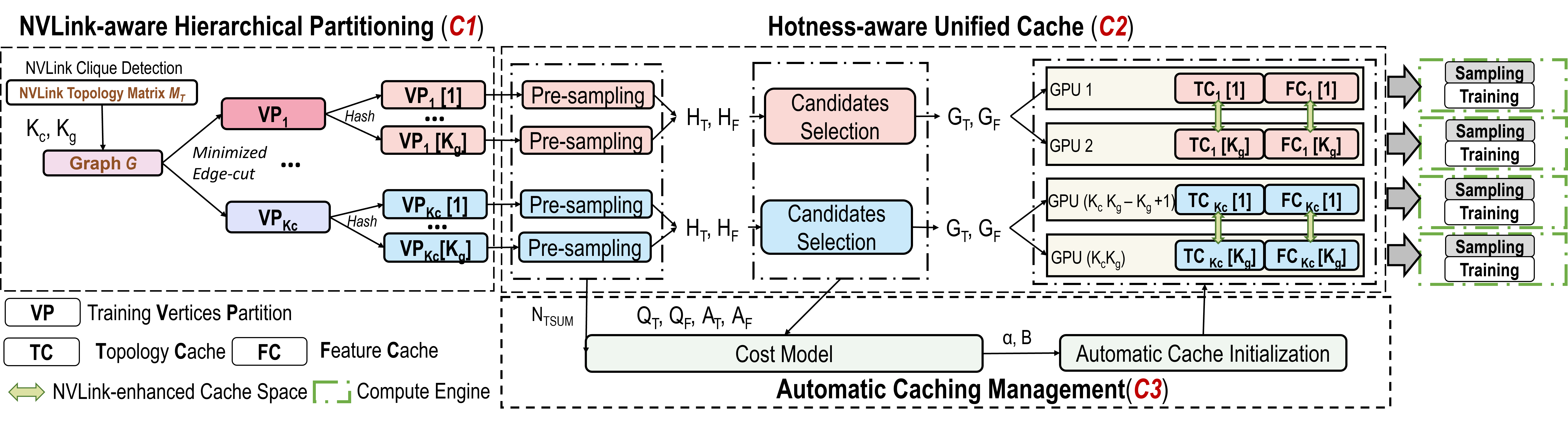} 
	\vspace{-5ex}	
	\caption{Design overview of \SystemName{}.  \SystemName{} consists of three main contributions C1, C2, and C3. 
    } 
    \vspace{-3ex}	
	\label{sysframework} 
\end{figure*}

In order to address the aforementioned performance issues of existing cache-based GNN systems, we propose \SystemName{}, a cache-optimal GNN system that can push the envelope of the multi-GPU system automatically for billion-scale GNN training. The overall design of ~\SystemName{} is presented in Figure~\ref{sysframework}. We propose an NVLink-aware hierarchical partitioning technique (Section~\ref{hiclustering}) in \SystemName{} that facilitates scaling up the cache capacity and reducing cache duplication in multi-GPU servers. To utilize GPU cache for both graph sampling and feature extraction, we present a hotness-aware unified cache (Section~\ref{unifiedcachetxt}) that maintains both the topology and feature caches to optimize the overhead of PCIe traffic. We also develop an automatic cache management mechanism (Section~\ref{automaticcaching}) to automatically
decide the memory allocations for both topology and feature caches. 


\subsection{NVLink-aware Hierarchical Partitioning}
\label{hiclustering}
Motivated by \textbf{observation O1}, we propose a simple yet effective graph partitioning mechanism, referred to as \textbf{hierarchical partitioning}, to facilitate cache scalability in ~\SystemName{}. 
Different from conventional graph partitioning algorithms which partition all edges/vertices of a graph into multiple tablets, hierarchical partitioning in \SystemName{} aims to divide the training vertices/edges into multiple disjoint tablets. The inputs of hierarchical partitioning are an NVLink topology matrix $M_{T}$ of the underlying multi-GPU server and a graph $G$. The output is an assignment plan disseminating training vertices/edges among GPUs. Specifically, the process of hierarchical partitioning mainly consists of four steps:

\noindent\textbf{S1: NVLink Clique Detection. }
With the topology matrix $M_{T}$ of the server, \SystemName{} employs a MaxCliqueDyn algorithm~\cite{maxcliquealgo} to identify the NVLink clique sets in $M_{T}$, and outputs the number of NVLink cliques $K_{c}$ and the number of GPUs in each clique $K_{g}$.

\noindent\textbf{S2: Inter-clique Graph Partitioning.}
To reduce the cache duplication between NVLink cliques, \SystemName{} uses an edge-cut minimizing partitioning algorithm, e.g., METIS~\cite{karypis1997metis} and XtraPulp~\cite{slota2017partitioning}, to split the input graph $G$ into $K_{c}$ partitions, i.e., $P_1$, $P_2$, ..., $P_{Kc}$, such that nodes are balanced among partitions and inter-partition edge-cuts are minimized. The training vertex set in $P_i$ is denoted as $VP_i$. As the training vertices are randomly selected from $G$, the training vertex sets of different partitions are almost of the equal size. The number of partitions is equal to the number of detected NVLink cliques, and each NVLink clique hosts the cache for a dedicated partition. 
This way, \SystemName{} can reduce the cache duplication between NVLink cliques and take advantage of cache locality within an NVLink clique.

\noindent\textbf{S3: Intra-clique Training Vertex Partitioning.} 
As GPUs within an NVLink clique can access each other's memory via low-latency high-throughput NVLink interconnect, hierarchical partitioning further hashes the training vertex set of each partition into $K_{g}$ tablets, where $K_{g}$ is the GPU number in a clique. E.g., $VP_i$ is split into $VP_i[1]$ and $VP_i[2]$ if $K_{g}$ equals 2. Each tablet is exclusively mapped to a GPU in the corresponding NVLink clique. We explain how to generate the cache for each training vertex tablet in Section~\ref{unifiedcachetxt}.

\noindent{\textbf{S4: Training Vertex Assignment.}
Finally, \SystemName{} assigns training vertices of each tablet to a corresponding GPU as the batch seeds, which will then be shuffled locally to generate mini-batches for graph sampling and training.}

As such, \SystemName{} provides better cache scalability and load balancing compared to existing systems. Figure~\ref{fig_comparison_of_four_types} shows the cache performance of \SystemName{} improves with the increase of GPUs almost linearly. Figure~\ref{loadbalance} illustrates that \SystemName{} has smaller fluctuations in the cache hit rates on multi-GPU servers with NVLink cliques of various sizes.

\subsection{Hotness-aware Unified Cache}
\label{unifiedcachetxt}
Motivated by the {\bf observation O2}, we propose a hotness-aware unified cache to cache both graph topology and graph features. In this Section, we introduce the detailed mechanism of the unified cache.

\vspace{-2ex}
\subsubsection{Cache Structure}
 The unified cache consists of two parts: the topology cache and the feature cache.
 In particular, the topology cache maintains out-edge neighbor IDs for each selected hot vertex in the format of a compressed sparse row (CSR). As for the feature cache, \SystemName{} stores the feature vectors of selected hot vertices in the format of a 2D array, where each row is the feature vector of a selected hot vertex. Note that, the selected vertices in the topology and feature caches could be different.

 \vspace{-2ex}
\subsubsection{Cache Construction}
\label{presamplingtxt}
 The construction of the unified cache is divided into three steps: (1) pre-sampling, (2) cache candidate selection, and (3) cache initialization. All the GPUs/NVLink cliques perform these steps concurrently to construct their own unified cache.



\begin{figure}[t]
        \centering	\includegraphics[width=3.2in]{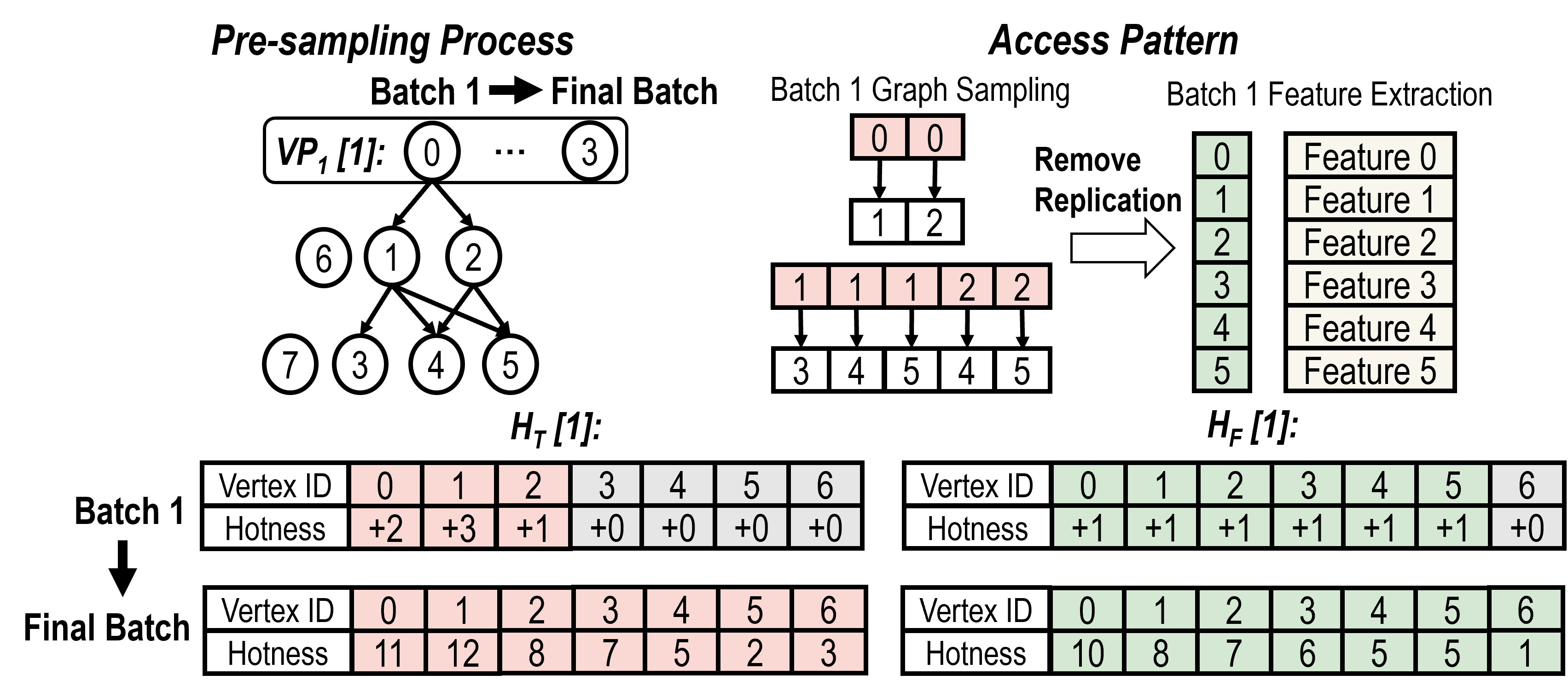} 
	\vspace{-2ex}	
	\caption{Update the hotness matrices of graph topology and features by pre-sampling. For simplicity, we only show the result for GPU 1.} 
	\vspace{-3ex}
	\label{gnnhotness} 
\end{figure}

\noindent\textbf{S1: Pre-sampling. }
Similar to GNNLab~\cite{yang2022gnnlab}, \SystemName{} adopts a pre-sampling phase\footnote{During pre-sampling, graph topology is stored in the CPU memory.} to estimate the hotness metrics of graph topology and feature data during the training phase. Once the process of hierarchical partitioning is completed, the training vertex tablet assigned to each GPU is determined, which is used as the input for pre-sampling. The output of pre-sampling includes two hotness matrices: topology hotness matrix $H_{T}$ and feature hotness matrix $H_{F}$. Each matrix's row represents the GPU IDs within an NVLink clique, the column represents the vertex IDs, and the element $H_{ij}$ of either matrix represents the hotness of the $j$-th vertex in the $i$-th GPU. 
During the pre-sampling, each GPU conducts a local shuffle on its own training vertex tablet to generate seeds for mini-batches, performs graph sampling for each mini-batch, and updates the corresponding row in $H_{T}$ and $H_{F}$. Figure~\ref{gnnhotness} shows a pre-sampling example. For $H_{T}$, whenever an edge is traversed during sampling, the hotness of its source vertex is incremented by $1$. For $H_{F}$, the hotness for each vertex that appears in the sample results of the mini-batch is incremented by 1. Additionally, \SystemName{} uses Intel® Performance Counter Monitor (PCM) ~\cite{pcmgit} to collect the summation of PCIe transactions number, $N_{TSUM}$, generated by all GPUs in an NVLink clique during pre-sampling.\footnote{$N_{TSUM}$ is further used by cost model's evaluation.}

\noindent\textbf{S2: Cache Candidate Selection. }
The objective of cache candidate selection is to select and disseminate the hot topology sub-structures and features among GPUs within the same NVLink clique based on pre-sampled hotness matrices. Thus this phase is conducted in the unit of NVLink clique, and each clique requires one GPU to perform the computation. The detailed process of cache candidate selection is presented in Algorithm~\ref{cslp}. 
In brief, this algorithm computes the global topology/feature hotness of all vertices, i.e., $A_T$ and $A_F$, in the NVLink clique by conducting a column-wise sum on $H_T$ and $H_F$, respectively (Line 1). $A_T$ and $A_F$ are then sorted in descending order to generate $Q_T$ and $Q_F$(Line 2). 
Next, We iterate $Q_T$ and $Q_F$ in order and assign every visited vertex to the GPU with the highest local hotness in $H_T$ and $H_F$.
For each GPU, we maintain two queues, i.e., $G_{T}$, $G_{F}$, whose order represents the priority of vertices to be included in this GPU cache. The outputs of Algorithm~\ref{cslp} are further used by the cost model (see Section~\ref{automaticcaching}) to generate the physical cache plan.

\noindent\textbf{S3: Cache Initialization and Fill-up. } \SystemName{}'s cache management automatically decides the cache ratio for topology and feature so that the overall throughput is maximized (see Section~\ref{automaticcaching}). Guided by this mechanism, \SystemName{} allocates memory for both the topology and feature cache ($TC$ and $FC$) of each GPU, and fetches the corresponding topology and feature data from CPU memory to fill up each GPU cache according to the corresponding cache orders in $G_{T}$ and $G_{F}$.

\begin{algorithm} [t]
	\SetAlFnt{\scriptsize} \linespread{1.0} \selectfont \caption{\sc Complete Sharing with Local Preference (CSLP)}
	\label{cslp}
	\SetKwInOut{Define}{Define}
	\newcommand\mycommfont[1]{\footnotesize\textcolor{gray}{#1}} 
    \SetCommentSty{mycommfont}
	
	\begin{footnotesize}
	      \SetKwInOut{Input}{Input}\SetKwInOut{Output}{Output}
       \SetKwInOut{Data}{Intermediate}
		\Input{
            $K_{g}$: number of GPUs per NVLink clique \\
            $H_F$: feature hotness matrix \\
            $H_T$: topology hotness matrix \\
            
            }
            \Output{
            ${A_F}$: accumulated vertex-wise feature hotness vector\\
            ${A_T}$: accumulated vertex-wise topology hotness vector\\
            ${Q_{T}}$: vertex ID queue representing clique-level topology order,\\
            ${Q_{F}}$: vertex ID queue representing clique-level feature order\\
            ${G_{T}}$: vertex ID queue representing GPU-level topology order\\
            ${G_{F}}$: vertex ID queue representing GPU-level feature order
            }

		
        
        \tcc{Step 1: Accumulate each vertex's hotness from $K_{g}$ GPUs. }
            ${A_F}$ = $H_F.columnWiseSum()$; 
            ${A_T}$ = $H_T.columnWiseSum()$;  
            
        \tcc{Step 2: Sort vertices in ${A_F}$ and ${A_T}$ }
            ${Q_{F}}$ <- SortbyKeyDescend(${A_F}$); 
            ${Q_{T}}$ <- SortbyKeyDescend(${A_T}$); 
            
        \tcc{Step 3: Assign each vertex to the GPU with the highest local hotness.}
		  \For{$v\_id$ in ${Q_{T}}$}{
            
          ${gpu\_id}$ = $max(H_{T}[1:K_{g}][v\_id]).index$; \\
          $G_{T}[gpu\_id].push(v\_id)$;
            
		}
  		\For{$v\_id$ in ${Q_{F}}$}{
            
          ${gpu\_id}$ = $max(H_{F}[1:K_{g}][v\_id]).index$; \\
          $G_{F}[gpu\_id].push(v\_id)$;
            
		}
    \end{footnotesize}
\end{algorithm}

\subsection{Automatic Cache Management}
\label{automaticcaching}
The design of the unified cache poses a new challenge: how to properly specify the cache size for graph topology and features under the constraint of GPU memory such that the overall training throughput is maximized.
 
The general idea is to predict the overall throughput under different cache plans and search for the best cache plan that maximizes overall throughput. We define the cache plan as a cache memory management setting ($B$, $\alpha$) at the NVLink clique granularity, where $B$ is the multi-GPU cache memory size in an NVLink clique and $\alpha$ is the memory ratio for topology cache. 
$B$ is identical among NVLink cliques and is by default set as the total multi-GPU memory minus the size of GPU memory reserved for GNN models and intermediate buffers in an NVLink clique. 
We need three steps to determine the optimal cache memory management setting ($B$, $\alpha$), 
as discussed in Sections~\ref{q1sec},~\ref{q2sec}, and~\ref{q3sec}. 

\subsubsection{Estimating Overall Throughput}
\label{q1sec}
The key goal of this Section is to build the relationship between the overall throughput and a cache plan. 
We build the relationship by estimating a key factor: the total PCIe traffic $N_{total}$, due to two reasons. First, the PCIe traffic is the major bottleneck of the overall system throughput, and lower PCIe traffic leads to higher overall system throughput. Second, varying cache plans major results in the variance of PCIe traffic.\footnote{Though NVLink traffic is also influenced by the cache plan, we neglect it since NVLink has a much higher bandwidth than PCIe. }
Because each NVLink clique maintains caches for its own partition, we independently select the optimal cache plan for each NVLink clique so as to minimize the PCIe traffic of each NVLink clique. Thus, the overall system's PCIe traffic is minimized. 

\subsubsection{ Cost Model to Estimate $N_{total}$}
\label{q2sec}


The key goal of this Section is to present a cost model to estimate $N_{total}$ under a specific cache plan ($B$, $\alpha$). First, given a specific cache plan ($B$, $\alpha$), we can calculate the topology cache size $m_{T}$ and the feature cache size $m_{F}$. Second, we find which vertices' topology/features should be stored in the topology/feature cache. Third, we estimate the PCIe traffic for graph sampling ($N_{T}$) and for feature extraction ($N_{F}$) with the current topology/feature cache utilization. At last, we estimate $N_{total}$ by adding up $N_{T}$ and $N_{F}$, as shown in Equation~\ref{overallcost}. 

\begin{equation}
    \begin{aligned}
    \label{overallcost}
    N_{total} &= N_{T} + N_{F}
    \end{aligned}
\end{equation}

To estimate $N_{T}$ and $N_{F}$, we need to collect other information apart from a given cache plan: the hotness vectors $A_{T}$ and $A_{F}$, the summation of PCIe transaction number $N_{TSUM}$ incurred by graph sampling, and the order queues of topology/feature cache candidates, $Q_{T}$ and $Q_{F}$. 


\noindent{\textbf{Estimating $N_{T}$.}
We estimate $N_{T}$ when the memory size of a topology cache under one specific cache plan ($B$, $\alpha$) is  $m_{T}$, where $m_{T} = B \times \alpha$. The estimation consists of three steps.
}

First, with $m_{T}$, we decide which vertices' topology should be cached. We define ${V}$ as the set of all vertices in the graph. And we define ${V_{Tcache}}$ as the set of all vertices whose topology is cached under current topology cache size $m_{T}$.
To get ${V_{Tcache}}$, we increase vertices and their topology into the cache with the growth of occupied topology cache memory by the order $Q_{T}$. Until the overall occupied topology cache memory size reaches $m_{T}$, we record ${V_{Tcache}}$. Equation~\ref{evalnt1} illustrates the relation between $m_{T}$ and ${V_{Tcache}}$, where $nc(v)$ means the neighbor count of the vertex $v$. Here we assume the data types are Uint64 and Uint32 for the row and the column indices of the compressed sparse row format (CSR), respectively. We use $s_{uint64}$ and $s_{uint32}$ to denote the number of bytes to store a single Uint64 and Uint32 data accordingly. 
\begin{equation}
\label{evalnt1}
    \sum_{v\in{V_{Tcache}}} (nc(v) \times s_{uint32} + s_{uint64}) = m_{T}
\end{equation}

Second, once we get ${V_{Tcache}}$, we can calculate the ratio of the PCIe transaction reduced by the topology cache by Equation~\ref{evalnt2}. Let $a_{T}(v)$ mean the topology hotness of a specific vertex $v$ ($a_{T}(v) \in A_{T}$). 

\begin{equation}
\label{evalnt2}
   R_{T} = \frac{ \sum_{v\in{V_{Tcache}}} a_{T}(v) }{\sum_{v\in{V}} a_{T}(v)})
\end{equation}

Third, we get $N_{T}$ by multiplying the entire PCIe transaction $N_{TSUM}$ with the ratio of PCIe transactions that can \textbf{not} be reduced by the topology cache.  
We can get $N_T$ by Equation~\ref{evalnt3}.
\begin{equation}
\label{evalnt3}
   N_{T} = N_{TSUM} \times (1 - R_{T})
\end{equation}


\noindent{\textbf{Estimating $N_{F}$.}
We explain how to calculate $N_{F}$ when the feature cache memory size is $m_{F}$, where $m_{F} = B \times (1 - \alpha)$. There are also three steps in estimation.
}

First, given $m_{F}$, we decide which vertices' features should be cached. We define ${V_{Fcache}}$ as the set of vertices whose feature data is cached.
Then we increase the vertices with their feature into cache by the order $Q_{F}$ until the occupied feature cache memory size reaches $m_{F}$, as defined in Equation~\ref{evalnf1}. $D$ represents the dimension of a feature vector and the feature data is the Float32 type each of which needs $s_{float32}$ bytes to store.
\begin{equation}
\label{evalnf1}
   \sum_{v\in{V_{Fcache}}} D\times s_{float32} =  m_{F}
\end{equation}

Second, as shown in Equation~\ref{evalnf2}, we calculate the total number of features $U_{F}$ that still needs transferring through PCIe with a feature cache. 

\begin{equation}
\label{evalnf2}
    U_{F} = \sum_{v\in{V}} (a_{F}(v)) - \sum_{v\in{V_{Fcache}}} (a_{F}(v))
\end{equation}

Third, we get $N_{F}$ by multiplying the transaction number needed by transferring one vertex's feature with the total number of features to be transferred, $U_{F}$, as shown in Equation~\ref{evalnf3}. 
Here $CLS$ means the transferred cache line size. $CLS$ might be different for various CPUs and GPUs. We can get the $CLS$ from PCM. E.g., $CLS$ equals 64 in our machine settings. And $a_{F}(v)$ means the hotness of a specific vertex $v$ ($a_{F}(v) \in A_{F}$).

\begin{equation}
\label{evalnf3}
   N_{F} = (\lceil \frac{D\times s_{float32}}{CLS} \rceil)\times U_{F}
\end{equation}

\subsubsection{Searching for Optimal Cache Plan in Parallel}
\label{q3sec}

The key goal of this Section is to efficiently determine the optimal cache plan for each clique. 
As discussed in Section~\ref{q1sec}, we search for the optimal cache plan independently with one GPU for each NVLink clique. In each NVLink clique, we first need to traverse $\alpha$ from 0 to 1 by an interval $\Delta \alpha$~\footnote{$\Delta \alpha$ is set to be $0.01$ by default.} to generate the candidate cache plans, and the calculate $N_{total}$ accordingly. Then we need to search $N_{total}$ sequences and find the smallest one with the dedicated $\alpha$. To minimize overhead, the process is well parallelized, including four steps:

First, we generate all the candidate cache plans in parallel and get sequences of $m_{T}$ and $m_{F}$ in each setting.

Second, we get the boundaries of cached vertices set ${V_{Tcache}}$ and ${V_{Fcache}}$ using Equations~\ref{evalnt1} and~\ref{evalnf1}, where the boundaries are the largest cached vertices' indexes in $Q_{T}$ and $Q_{F}$. To do so, 
we get the topology and feature memory size of every single vertex in parallel and store them in two arrays, $S_{Tsingle}$ and $S_{Fsingle}$, following the vertices order, $Q_{T}$ and $Q_{F}$. Next, we calculate the cumulative sum of $S_{Tsingle}$ and $S_{Fsingle}$ by a parallel inclusive scan and get $S_{Tsum}$ and $S_{Fsum}$.
Then for each cache plan with $m_{T}$ and $m_{F}$, we use a parallel binary search towards $S_{Tsum}$ and $S_{Fsum}$ to get the boundary indexes of vertices, respectively.

Third, we get the $R_{T}$ and $U_{F}$ according to Equations~\ref{evalnt2} and~\ref{evalnf2}. To do so, we calculate the cumulative sum of $A_{T}$ and $A_{F}$ by a parallel inclusive scan and get $A_{Tsum}$ and $A_{Fsum}$.
Then for each cache plan, we lookup $A_{Tsum}$ and $A_{Fsum}$ with the boundary indexes of vertices set ${V_{Tcache}}$, ${V_{Fcache}}$, and get $\sum_{v\in{V_{Tcache}}} a_{T}(v)$ and $\sum_{v\in{V_{Fcache}}} a_{F}(v)$, respectively. Similarly, after lookup the largest indexes in $A_{Tsum}$ and $A_{Fsum}$, we get $\sum_{v\in{V}} a_{F}(v)$ and $\sum_{v\in{V}} a_{F}(v)$. As such, we can get the corresponding $R_{T}$ and $U_{F}$.

At last, we calculate $N_{T}$ and $N_{F}$ for each cache plan according to Equation~\ref{evalnt3} and~\ref{evalnf3}.
Then we get $N_{total}$ by Equation~\ref{overallcost} and search in parallel for the smallest $N_{total}$ with the corresponding $\alpha$.

After getting the optimal cache plans ($B$, $\alpha$), \SystemName{} can automatically allocate the cache space and fill up the cache.


\vspace{-2ex}


\vspace{1ex}
\section{Implementation of \SystemName{}}
\label{finegrainedpipeline}
\vspace{-2ex}
\begin{figure}[t]
    \centering	\includegraphics[width=3.2in]{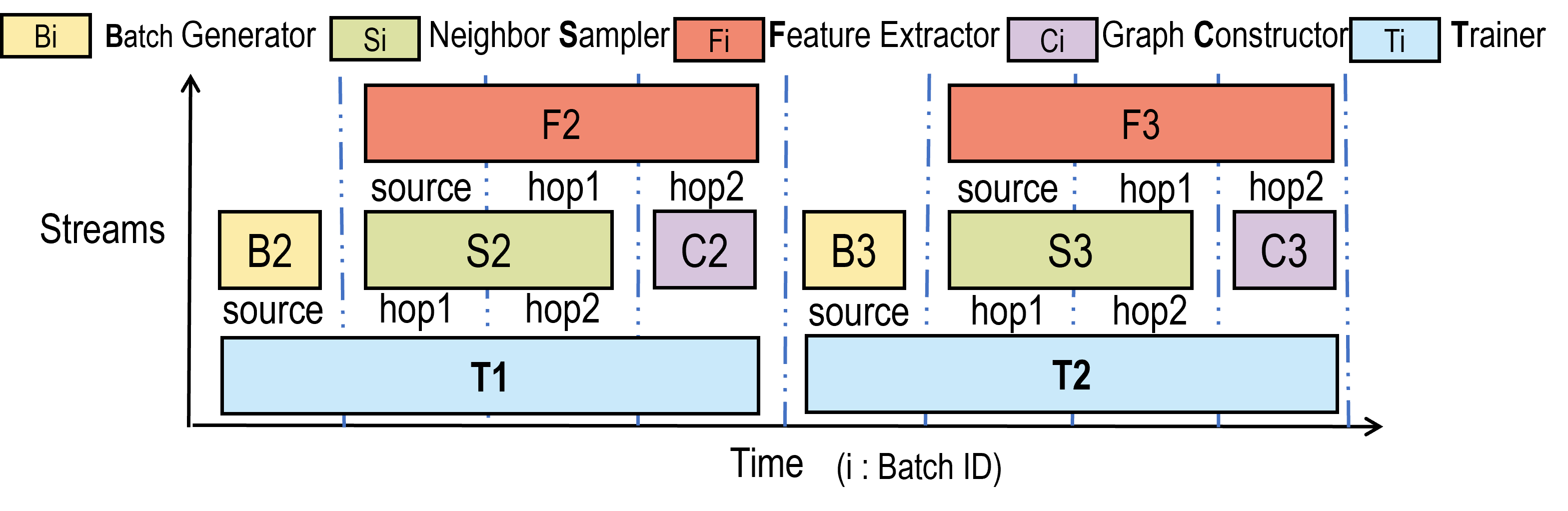} 
	\vspace{-2ex}	
	\caption{An example of fine-grained GNN training pipeline for 2-hop GraphSAGE model.} 
	\label{pipeline} 
\end{figure} 

\SystemName{} mainly consists of two components, which are the sampling server and the training backend. The sampling server is implemented from scratch and the training backend is built on top of Pytorch~\cite{paszke2019pytorch}. The sampling server is responsible for generating sampled results, and the training backend takes the sampled results as input to train the GNN models.

In \SystemName{}, every GPU executes the graph sampling, feature extraction, and model training stages, and all these stages are scheduled in a fine-grained pipeline to fully utilize the GPU computation cycles. Figure~\ref{pipeline} illustrates how the training process is pipelined for a 2-hop GraphSAGE~\cite{hamilton2017inductive} model. In order to improve the overall throughput, we design an inter-batch pipeline overlapping the tasks of the sampling server and the training backend for different batches. E.g., the training of batch $B_i$ can be overlapped with the sampling and feature extraction of batch $B_{i+1}$. To further improve the throughput of sampling and feature extraction, we design an intra-batch pipeline inside the sampling server. Specifically, 
we break down the workloads of the sampling server into four types, each of which corresponds to a type of operator: (1) Batch generator shuffles the local training vertices to generate seeds for mini batches; (2) Neighbor sampler executes the L-hop neighbor sampling; (4) Feature extractor extracts the feature of the batch seeds and vertices in the sampled results; (4) Graph constructor is used to generating the subgraph based on the sampled results. For the same batch, graph sampling and graph construction can be overlapped with feature extraction.
\vspace{-2ex}
\section{Evaluation}

\subsection{Experimental Setting}
\label{expsettingtxt}


		


\begin{figure*}[t]
    \centering
    \subfloat[DGX-V100, GraphSAGE]{
        \includegraphics[width=1.7in]{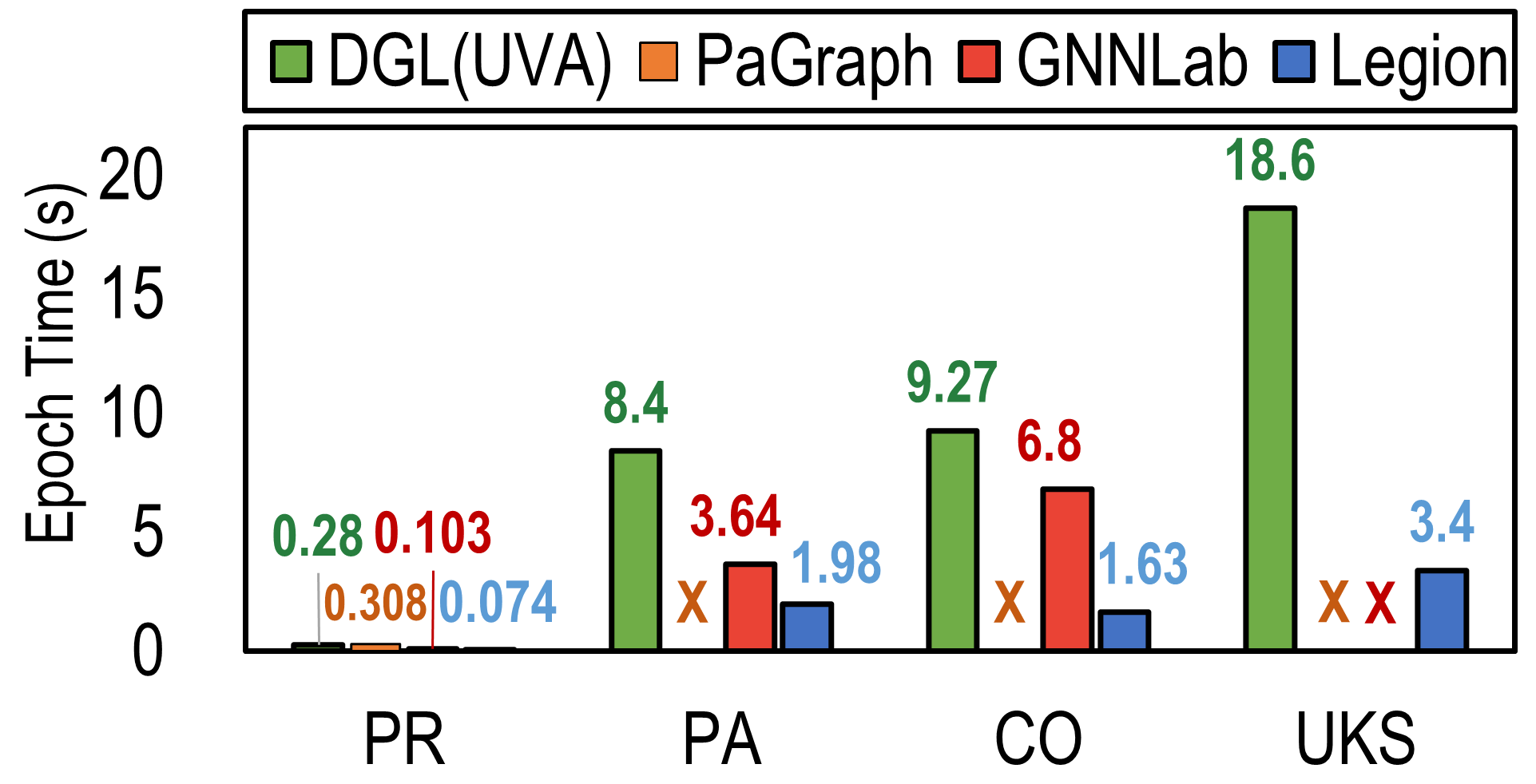}
        \label{end2endspeedupdgxgraphsage}
    }
    \subfloat[DGX-V100, GraphSAGE]{
	\includegraphics[width=1.7in]{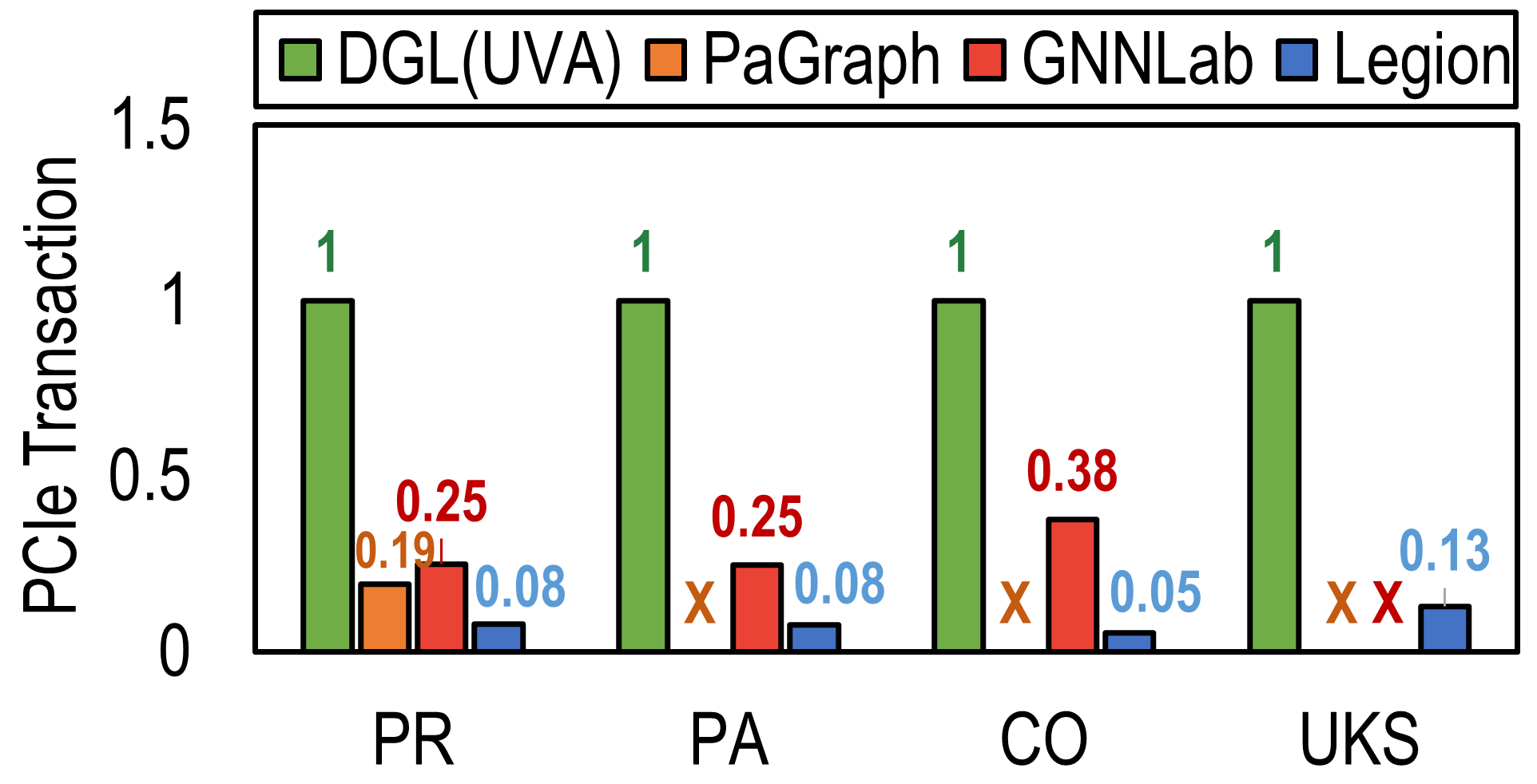}
        \label{end2endtransdgxgraphsage}
    }
    \subfloat[DGX-A100, GraphSAGE]{
    	\includegraphics[width=1.7in]{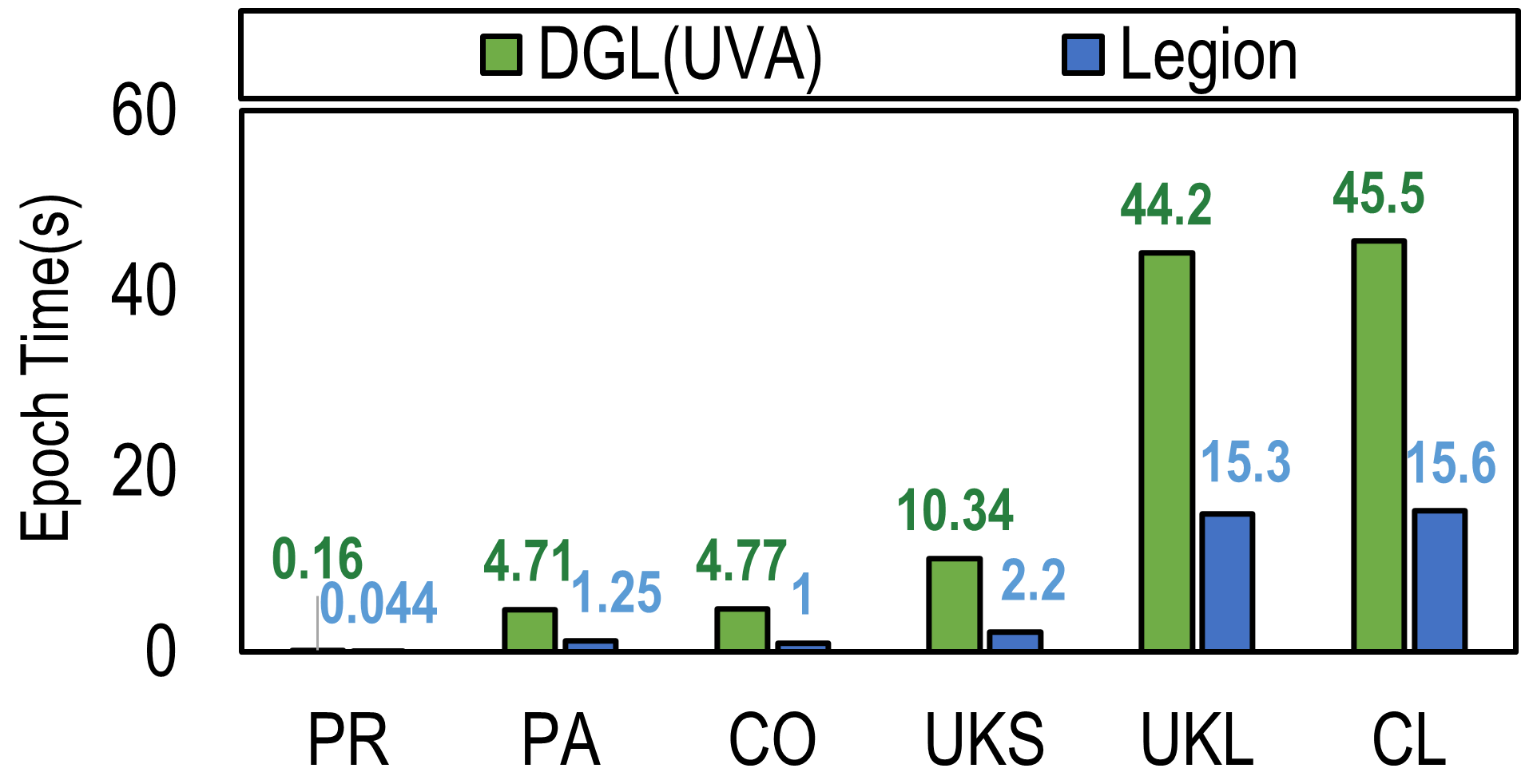}
        \label{end2endspeeduphgxgraphsage}
    }
    \subfloat[DGX-A100, GraphSAGE]{
	\includegraphics[width=1.7in]{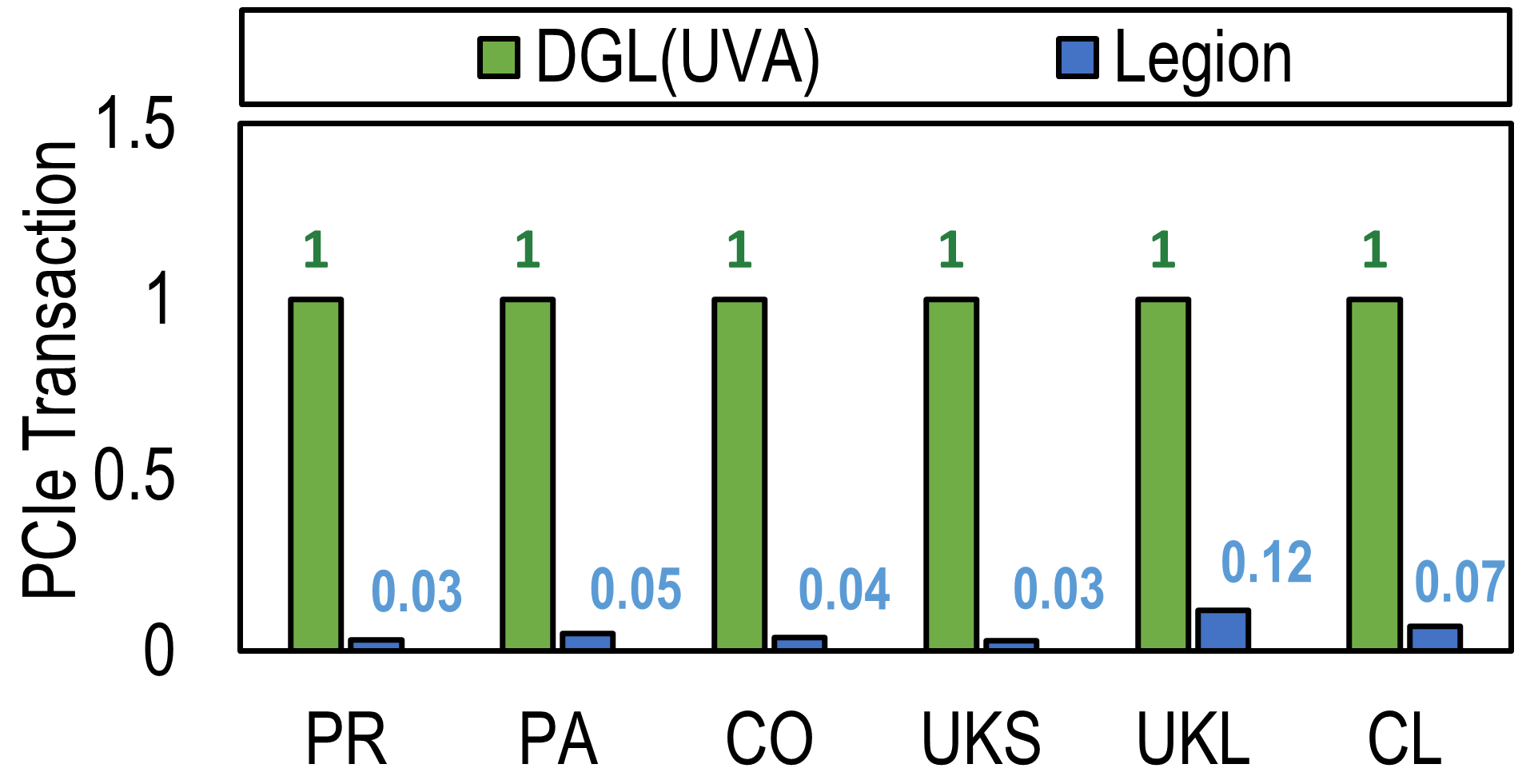}
        \label{end2endtranshgxgraphsage}
    }
    \\
    \subfloat[DGX-V100, GCN]{
        \includegraphics[width=1.7in]{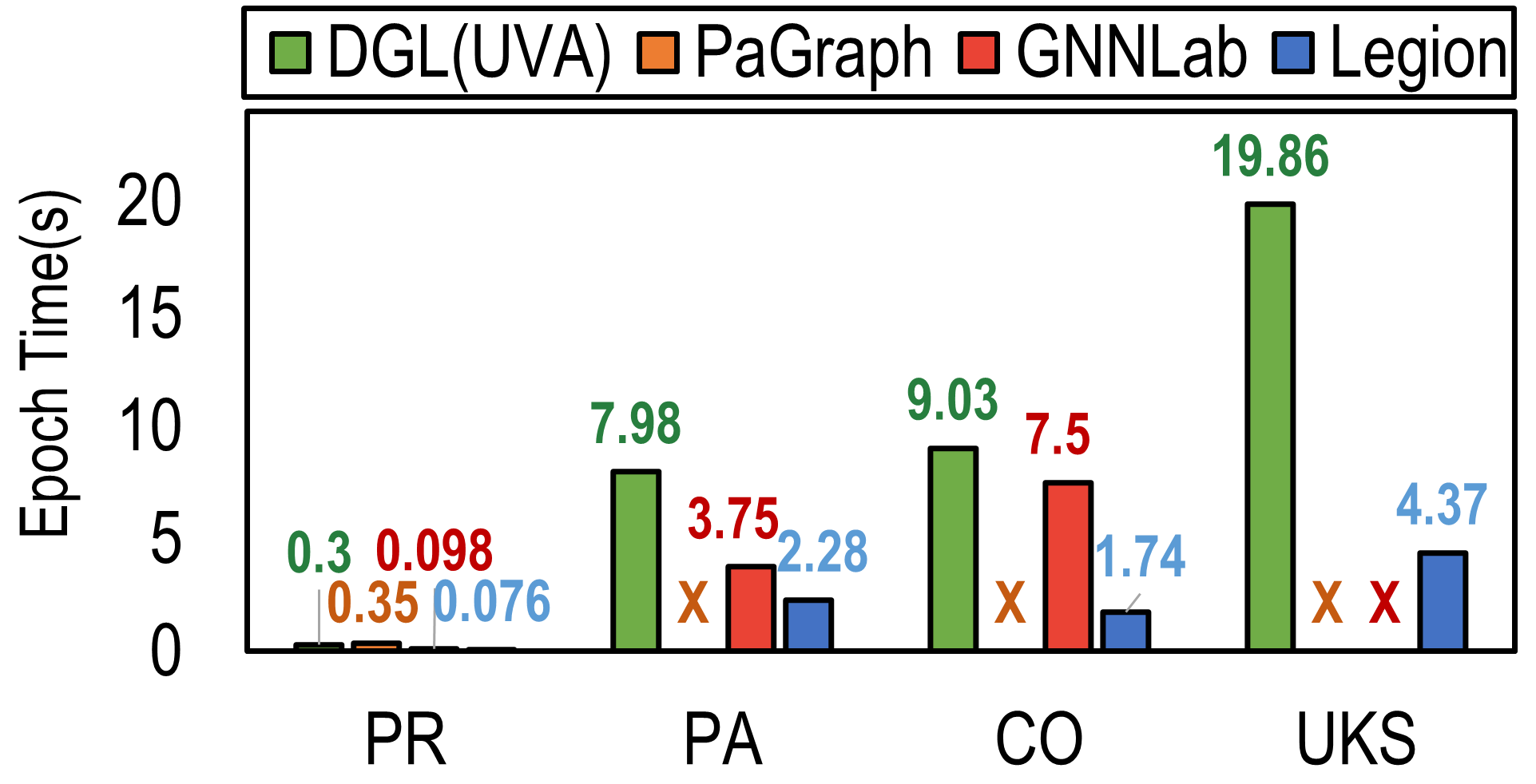}
        \label{end2endspeedupdgxgcn}
    }
    \subfloat[DGX-V100, GCN]{
	\includegraphics[width=1.7in]{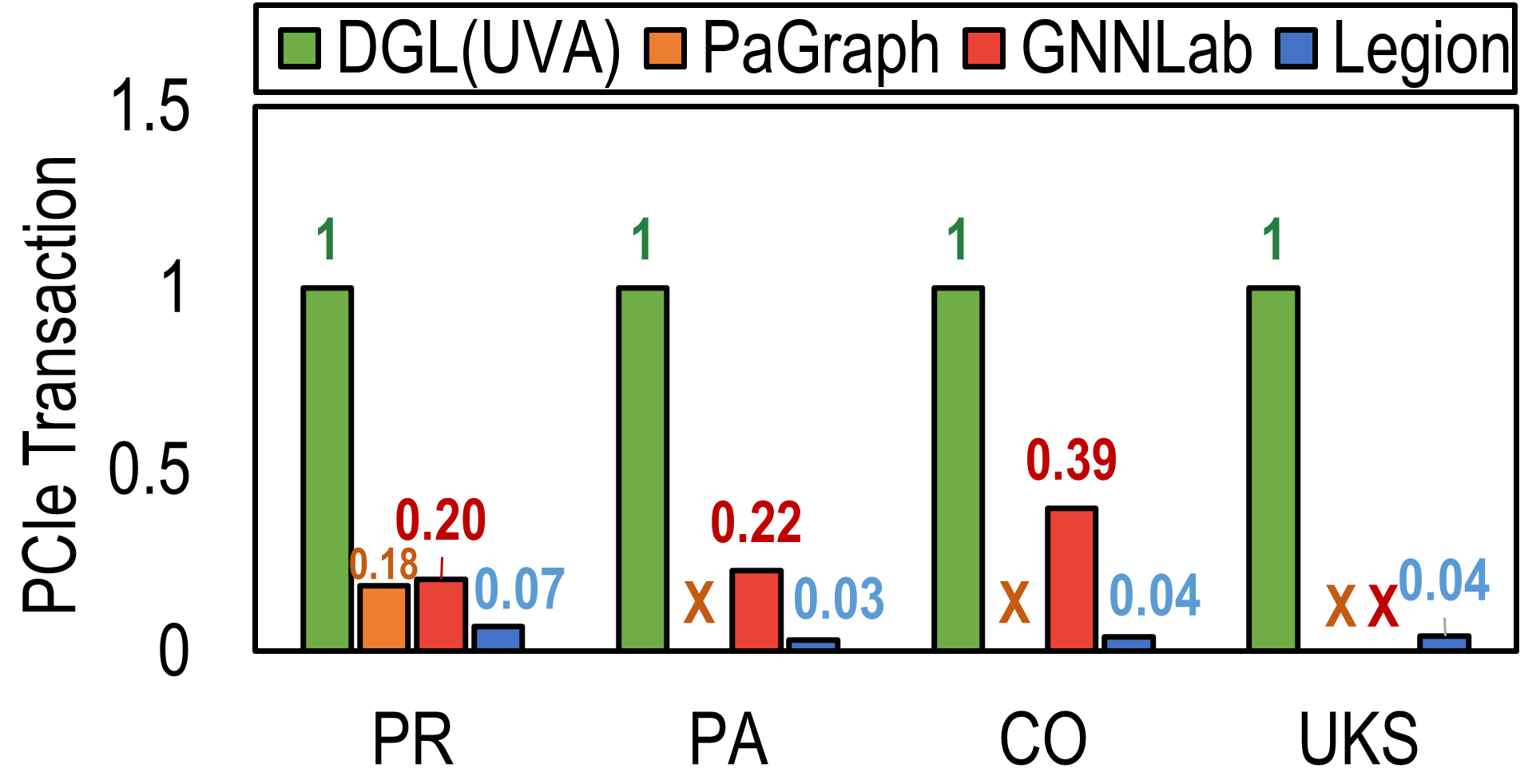}
        \label{end2endtransdgxgcn}
    }
    \subfloat[DGX-A100, GCN]{
    	\includegraphics[width=1.7in]{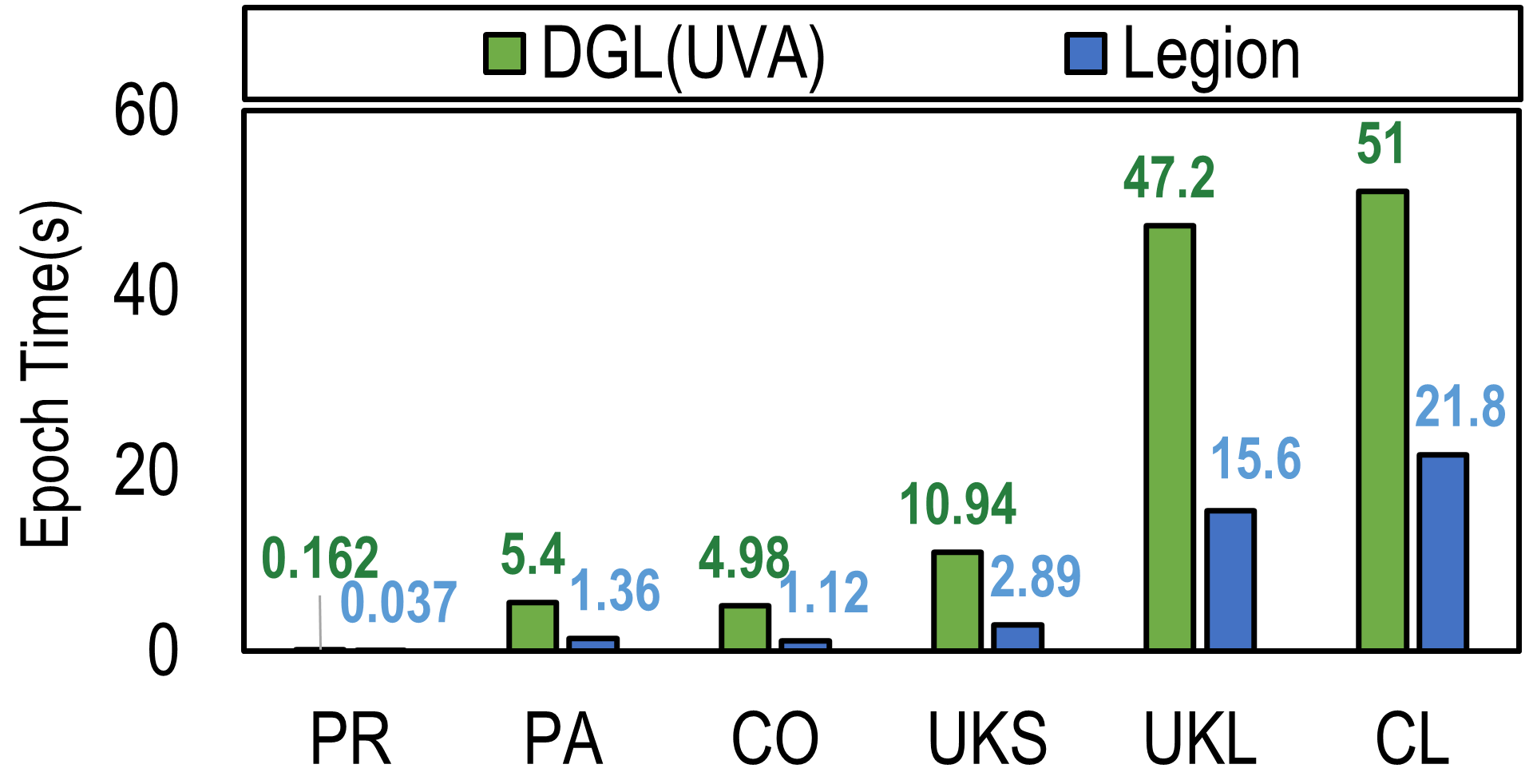}
        \label{end2endspeeduphgxgcn}
    }
    \subfloat[DGX-A100, GCN]{
	\includegraphics[width=1.7in]{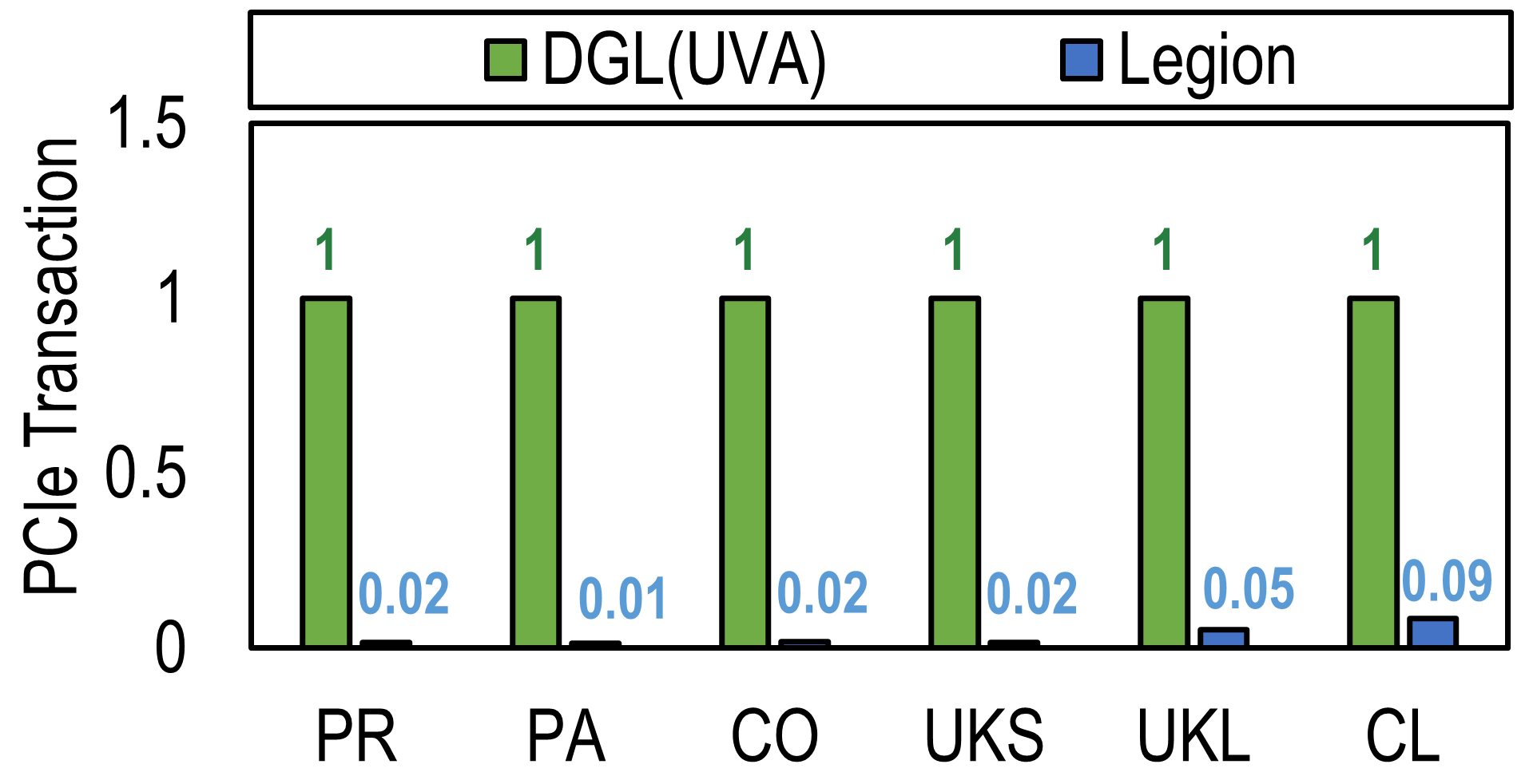}
        \label{end2endtranshgxgcn}
    }
    \vspace{-2ex}
    \caption{Overall performance of \SystemName{} comparing with state-of-the-art systems. ``$\times$'' denotes OOM (out of memory).}
    \vspace{-2ex}
    \label{end2endperformance}
\end{figure*}

\noindent{\bf Experimental Platform.} The experiments are conducted using three different GPU servers: DGX-V100, Siton, and DGX-A100, as shown in Table~\ref{serverstat}. For DGX-A100, we set the upper limit of GPU memory to 40 GB.

\begin{table} [t]
\renewcommand\arraystretch{1.4}
\setlength\tabcolsep{3pt}
	\centering

		\begin{scriptsize}
		
	\vspace{-0.5ex}
	\caption{GPU Server Statistics.}           \vspace{-0.5ex}
	\label{serverstat}
	\begin{tabular}{|c||c|c|c|}
		\hline
		{\bf Server} &  {\bf DGX-V100 }& {\bf Siton }&  {\bf DGX-A100 }\\
		\hline
            \hline
		{\bf GPU Type} & 16GB-V100x8 & 40GB-A100x8 & 80GB-A100x8 \\ 
		\hline
            {\bf NVLink Topo.} & $K_{c}=2$, $K_{g}=4$ & $K_{c}=4$, $K_{g}=2$ & $K_{c}=1$, $K_{g}=8$\\ 
		\hline
  		{\bf PCIe Gen.} & 3.0x16 & 4.0x16 & 4.0x16 \\ 
		\hline
              {\bf  PCIe Topo.} & \makecell[c]{ 4 switches,\\  2 GPUs/switch} & \makecell[c]{ 2 switches,\\ 4 GPUs/switch} & \makecell[c]{ 4 switches,\\  2 GPUs/switch} \\ 
		\hline
    	{\bf CPU Mem.} & 384GB & 1TB & 1TB \\ 
  		\hline
      	{\bf  CPU Core Num.} & \makecell[c]{ 96} & \makecell[c]{ 104} & \makecell[c]{ 128}	 \\ 
		\hline
      	{\bf  Sockets, NUMA Num.} &  2, 1 &  2, 2 &  2, 1 \\ 
		\hline
	\end{tabular}
	\vspace{-1ex}
		\end{scriptsize}
\end{table}


\noindent{\bf GNN Models.} We use two sampling-based GNN models: GraphSAGE~\cite{hamilton2017inductive} and GCN~\cite{kipf2016semi}, which both adopt a 2-hop random neighbor sampling. The sampling fan-outs are 25 and 10. The dimension of the hidden layers in both models is set to 256. Similar to existing work~\cite{yang2022gnnlab}, the batch size is set to 8000. Unless explicitly explained, node classification is used as the GNN task.

\noindent{\bf Datasets. } We conduct our experiments on multiple real-world graph datasets with various scales. Table~\ref{dataset} shows the dataset characteristics. The Products (PR) and Paper100M (PA) are available in Open Graph Benchmark~\cite{hu2020open}. The Com-Friendster (CO) graph is an online gaming network~\cite{yang2012defining}. And the Uk-Union (UKS), UK-2014 (UKL), and Clue-web (CL) are from WebGraph~\cite{BoVWFI, BRSLLP, BCSU3, BMSB}. As CO, UKS, UKL, and CL have no feature, we manually generate the features with the dimension specified as $128$ or $256$. Following PR's setting, we choose 10\% of vertices from each graph as training vertices. 

\begin{table} [t]
\renewcommand\arraystretch{1.4}
\setlength\tabcolsep{3pt}
	\centering

		\begin{scriptsize}
		
	\vspace{-1.5ex}
	\caption{Dataset Statistics.} \vspace{-0.5ex}
	\label{dataset}
	\begin{tabular}{|c||c|c|c|c|c|c|c|}
		\hline
		{\bf Dataset} &  {\bf PR}&  {\bf PA }  &  {\bf CO} &  {\bf UKS} & {\bf UKL} & {\bf CL } \\
		\hline
            \hline
		{\bf Vertices} &2.4M & 111M & 65M & 133M & 0.79B & 1B \\ 
		\hline
      	{\bf Edges} & 120M & 1.6B & 1.8B & 5.5B & 47.2B & 42.5B \\ 
		\hline
            {\bf Topology Storage} & 640M & 6.4GB & 7.2GB & 22GB & 189GB & 170GB \\ 
		\hline
  		{\bf Feature Size} & 100 & 128 & 256 & 256 & 128 & 128 \\ 
		\hline
    	{\bf Feature Storage} & 960M & 56GB & 65GB & 136GB & 400GB & 512GB \\ 
		\hline
	\end{tabular}
	\vspace{-2ex}

		\end{scriptsize}
\end{table}

\noindent{\bf Baselines. } We use DGL~\cite{wang2019deep}, PaGraph~\cite{lin2020pagraph} and GNNLab~\cite{yang2022gnnlab} as the baseline systems. The DGL version is v0.9.1, which supports accessing graph topology and features via the UVA technique. We don't compare with Quiver~\cite{torchquiver} in the overall performance experiment as its open-sourced version cannot support training on servers with 8 GPUs. Instead, we implement a Quiver-like multi-GPU cache mechanism in \SystemName{} for comparison in Section~\ref{experimentofhipartition}.

\subsection{End-to-end Performance}
We compare the end-to-end performance of \SystemName with baseline systems on the DGX-V100 and DGX-A100 servers. 
On the DGX-V100 server, we evaluate PR, PA, CO, and UKS graphs whose graph topology and features can fit into 384 GB CPU memory. On the DGX-A100 server, we evaluate all six graphs. As PaGraph and GNNLab are implemented using CUDA 10 which cannot support A100 GPU,  we exclude them from the experiments using DGX-A100.

\noindent{\textbf{Baseline Configuration. }For all the baselines, we manually adjust their configurations to achieve optimal performance. DGL uses the UVA mode, where sampling is performed in GPU, and the topology and features are all stored in CPU memory. The number of worker threads in PaGraph is set to be $64$ to maximize the CPU sampling throughput. For GNNLab, we adjust the numbers of sampling and training GPUs such that the overall throughput is maximized. In contrast, \SystemName{} relies on its automatic cache management mechanism to generate the unified cache plan.
}

\noindent{\textbf{Evaluation Metrics. }
We record the average epoch time for all systems.
We also use PCM~\cite{pcmgit} to measure the maximum PCIe counter value across different sockets and report the normalized values based on the result of DGL for all systems. 


\begin{figure*}[t]
    \centering
        \includegraphics[width=7in]{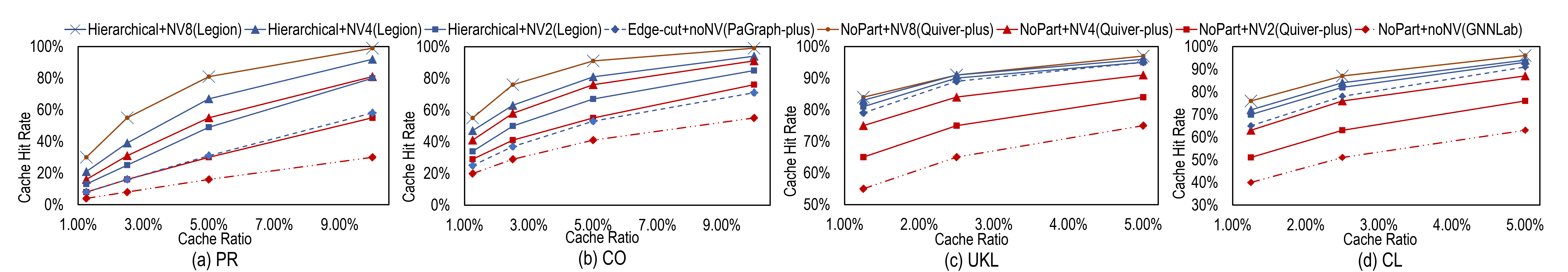}
        \label{cachehit}
    \vspace{-3ex}
    \caption{Effect of graph partition strategies (NoPart: no partitioning; Edge-cut: partitioning minimizing edge-cut; Hierarchical: hierarchical partitioning) to multi-GPU cache in terms of cache hit rate, with different NVLink infrastructures. (noNV: disable NVLinks; NV2: $K_{c}$ = 4 and $K_{g}$ = 2; NV4: $K_{c}$ = 2 and $K_{g}$ = 4; NV8: $K_{c}$ = 1 and $K_{g}$ = 8; ).}.
    \vspace{-4ex}
    \label{hiclusterexp}
\end{figure*}

\noindent{\textbf{Support training on large graphs. } 
As shown in Figures~\ref{end2endspeedupdgxgraphsage}, \ref{end2endspeedupdgxgcn}, \ref{end2endspeeduphgxgraphsage} and \ref{end2endspeeduphgxgcn}, \SystemName{} outperform all the baseline systems in every setting. Specifically, \SystemName{} achieves 3.78-5.69$\times$ speedup for GraphSAGE (3.5-5.19$\times$ for GCN) on DGX-V100 and 2.89-4.77$\times$ speedup for GraphSAGE (2.34-4.45$\times$ for GCN) on DGX-A100 over DGL(UVA). Figures~\ref{end2endtransdgxgraphsage}, \ref{end2endtransdgxgcn}, \ref{end2endtranshgxgraphsage} and \ref{end2endtranshgxgcn} show that, compared with the baselines, \SystemName{} can sufficiently utilize the multi-GPU cache to minimize PCIe traffic incurred by CPU-GPU data transferring. GNNLab runs out of GPU memory for UKS on DGX-V100 as the size of graph topology exceeds the capacity of single GPU memory. PaGraph runs out of the CPU memory for most graphs except for PR on DGX-V100, as the memory management in PaGraph incurs extra memory overheads, including duplicated multi-hop neighbors in CPU memory and redundant intermediate buffers generated during computation.}

\noindent{\textbf{Speedup over SOTA system on small graphs. }
\SystemName{} achieves 1.39-4.18$\times$ speedup for GraphSAGE (1.29-4.32$\times$ speedup for GCN) over GNNLab on the small graphs (PR, PA, CO). The performance gain mainly comes from two aspects.
First, Figure~\ref{end2endtransdgxgraphsage} and \ref{end2endtransdgxgcn} show that \SystemName{} significantly reduces the PCIe traffic for PA and CO, as it has a scalable multi-GPU cache design compared with GNNLab. The reduction of PCIe traffic relieves the CPU-GPU communication bottleneck such that the overall performance is improved. Second, \SystemName{} can use all GPUs for model training, while GNNLab needs to allocate several GPUs for sampling exclusively due to its factored design. In \SystemName{}, the graph sampling is overlapped by model training due to the fine-grained pipeline (see Section~\ref{finegrainedpipeline}).
E.g., when training GraphSAGE using the PR dataset, all the topology and feature data can be stored in GPU memory in both \SystemName{} and GNNLab. However, \SystemName can use 8 GPUs for training while GNNLab only uses 4 GPUs for training (see Figures~\ref{end2endspeedupdgxgraphsage}).

}

}

\vspace{-2ex}
\subsection{Effect of Hierarchical Partitioning}
\label{experimentofhipartition}
In this experiment, we examine the effect of hierarchical partitioning in \SystemName{}.
We report the cache hit rates under different partition strategies 
in all three GPU servers: DGX-V100 (NV4: $K_{c}$ = 2 and $K_{g}$ = 4), Siton (NV2: $K_{c}$ = 4 and $K_{g}$ = 2) and DGX-A100 (NV8: $K_{c}$ = 1 and $K_{g}$ = 8). 

\begin{figure}[t]
	\includegraphics[width=3.3in]{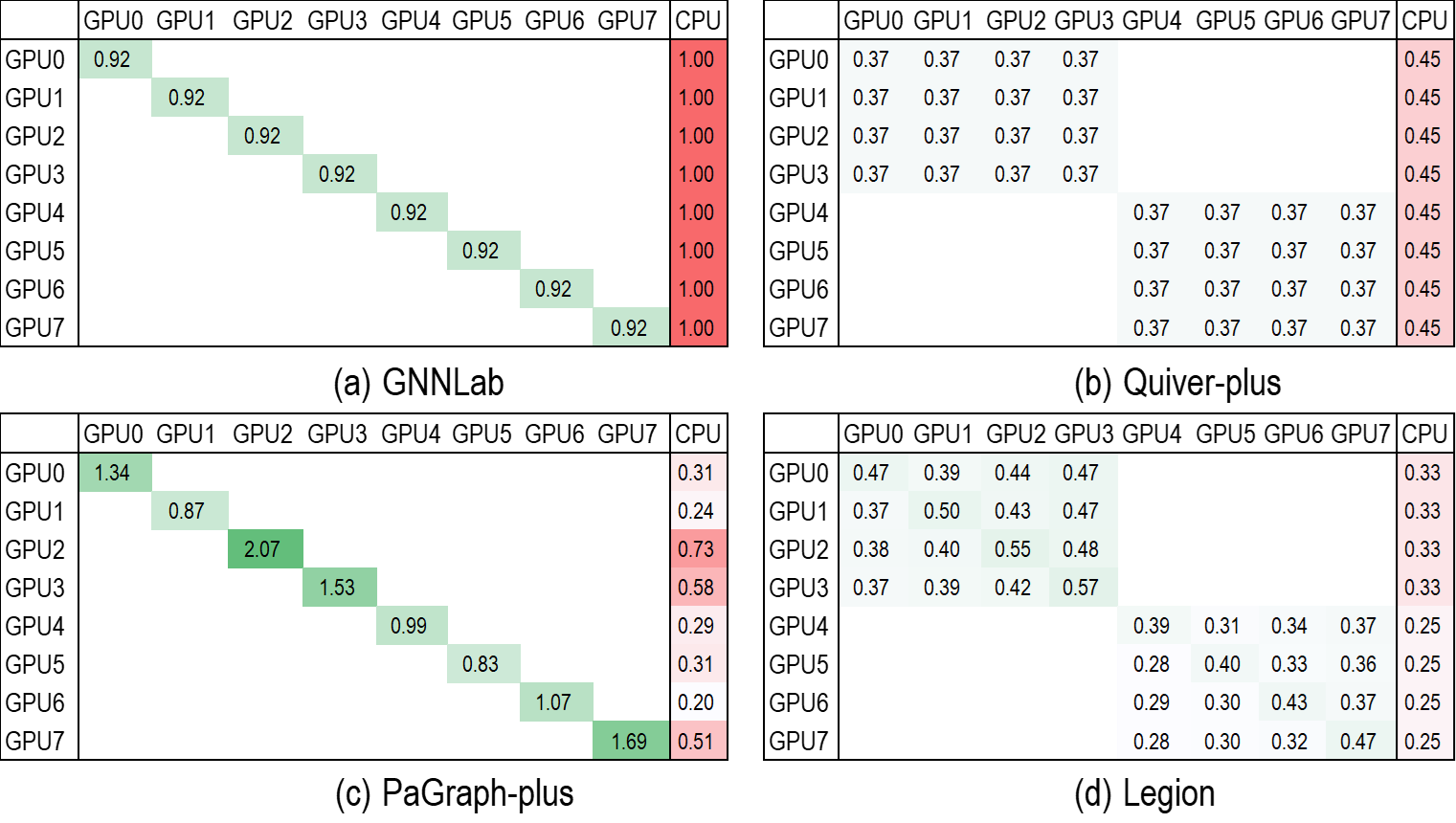} 
 	\vspace{-1ex}
	\caption{Data transferring in feature extraction of PA dataset on DGX-V100 (NV4). The rows and columns of each matrix denote the destination and source of data transferring. The right-most (red) column records the data transferring volume from CPU to GPU via PCIe. The middle (green) columns represent the GPU-GPU data transferring volume.  We normalize the recorded values based on the CPU-GPU data transferring volumes in GNNLab. } 
    \vspace{-3ex}
	\label{cachesharing} 
\end{figure} 

\subsubsection{Cache Performance}
\label{cacheperformanceexp}

\noindent{\textbf{Baselines. }For a fair comparison, we implement the cache designs of GNNLab, PaGraph-plus (described in Section~\ref{clusteringmotivation}), and Quiver-plus in \SystemName{} and compare their cache hit rates. Specifically, GNNLab maintains a globally replicated cache among all GPUs without using NVLinks (noPart+noNV). Quiver-plus enables NVLink and maintains replicated cache among NVLink cliques
(noPart+NV2 / noPart+NV4 / noPart+NV8). PaGraph-plus takes the XtraPulp~\cite{slota2017partitioning} partitioning which minimizes across-partition edge-cuts and disables NVLinks (Edge-cut+noNV). \SystemName{} uses hierarchical partitioning (inter-NVLink-clique partitioning: XtraPulp) and enables NVLink (Hierarchical+NV2 / Hierarchical+NV4 / Hierarchical+NV8). We use the pre-sampling hotness metric for all these cache designs. The in-degree-based hotness metric in the original PaGraph and Quiver design are replaced with the pre-sampling hotness metric in Pagraph-plus and Quiver-plus, which has a better performance on cache hit rates~\cite{yang2022gnnlab}.}

The datasets used in this experiment are PR, CO, UKL, and CL. We vary the cache ratio from 1.25\% $|V|$ to 10\% $|V|$ for PR and CO. For UKL and CL whose sizes are relatively large, the cache ratio varies from 1.25\% $|V|$ to 5\% $|V|$. Figure~\ref{hiclusterexp} shows that, for almost all the experiment settings, \SystemName{} has the highest cache hit rate. Specifically, \SystemName{} obviously outperforms Quiver-plus in the cases of NV2 and NV4, since \SystemName{} can reduce the inter-NVLink-clique cache duplication and achieves higher multi-GPU memory utilization compared with Quiver-plus. For the case of NV8, as all GPUs are in the same NVLink clique, the inter-clique graph partitioning in \SystemName{} can be skipped, and hierarchical partitioning turns into hash partitioning among all the GPUs, which is identical to Quiver-plus in the case of NV8. \SystemName{} outperforms PaGraph-plus because it has much less cache duplication. Specifically, PaGraph-plus's cache mechanism may replicate vertices with high global hotness on multiple GPUs. Compared with GNNLab, \SystemName{} has higher cache hit rates as it can scale up the cache capacity with the increase of GPUs, while GNNLab replicates the same feature cache across all GPUs.
These results demonstrate that \SystemName{} can effectively adapt the cache plan to optimize the cache performance for multi-GPU servers with various NVLink topologies.

\subsubsection{Data Transferring in Feature Extraction}
In this experiment, we demonstrate the GPU-GPU and CPU-GPU data transferring volume during feature extraction using the PA dataset. Specifically, we perform the graph sampling and feature extraction stages using the PA graph on DGX-V100 (NV4) and record the data transferring volumes of feature extraction on each GPU in the format of a traffic matrix. We use GNNLab, PaGraph-plus, and Quiver-plus as the baselines, and set the feature cache ratio on each GPU to 2.5\% $|V|$. The results are presented in Figure~\ref{cachesharing}. We can see that \SystemName{}'s data transferring volume from CPU to GPU is the smallest, indicating the best cache performance among the compared systems. As it is the GPU with the largest CPU-GPU data transferring volume that dominates the overall performance, although \SystemName{}'s CPU-GPU volumes on some GPUs are higher than PaGraph-plus, \SystemName{} can still outperform PaGraph-plus because its largest CPU-GPU volume is lower than that of PaGraph-plus.

\vspace{-3ex}
\subsubsection{Model Convergence}
\begin{figure}[t]
    \vspace{-1ex}
    \centering
    \subfloat[GraphSAGE]{
        \includegraphics[width=1.6in]{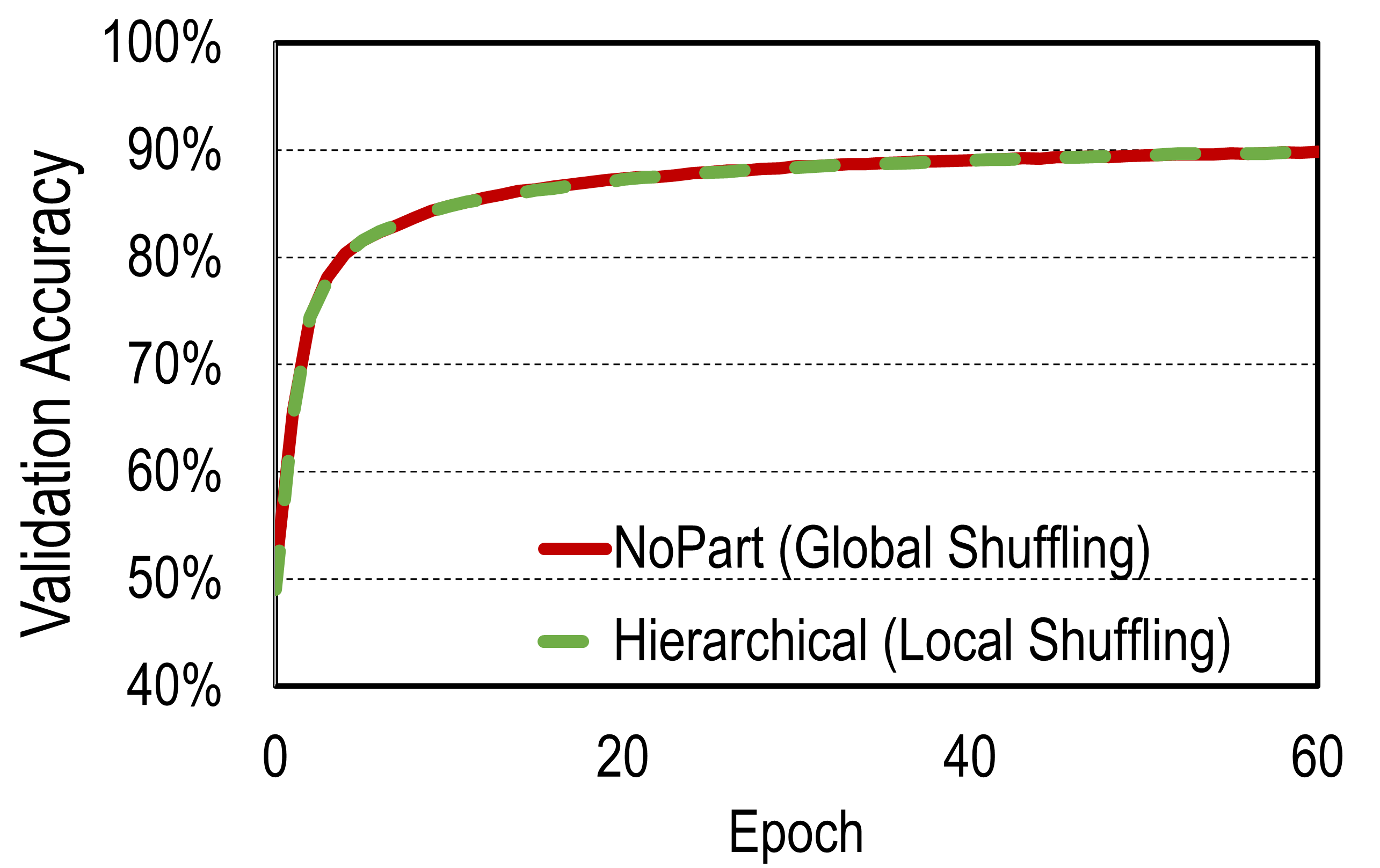}
        \label{convergencePRGraphsage}
    }
    \subfloat[GCN]{
	\includegraphics[width=1.6in]{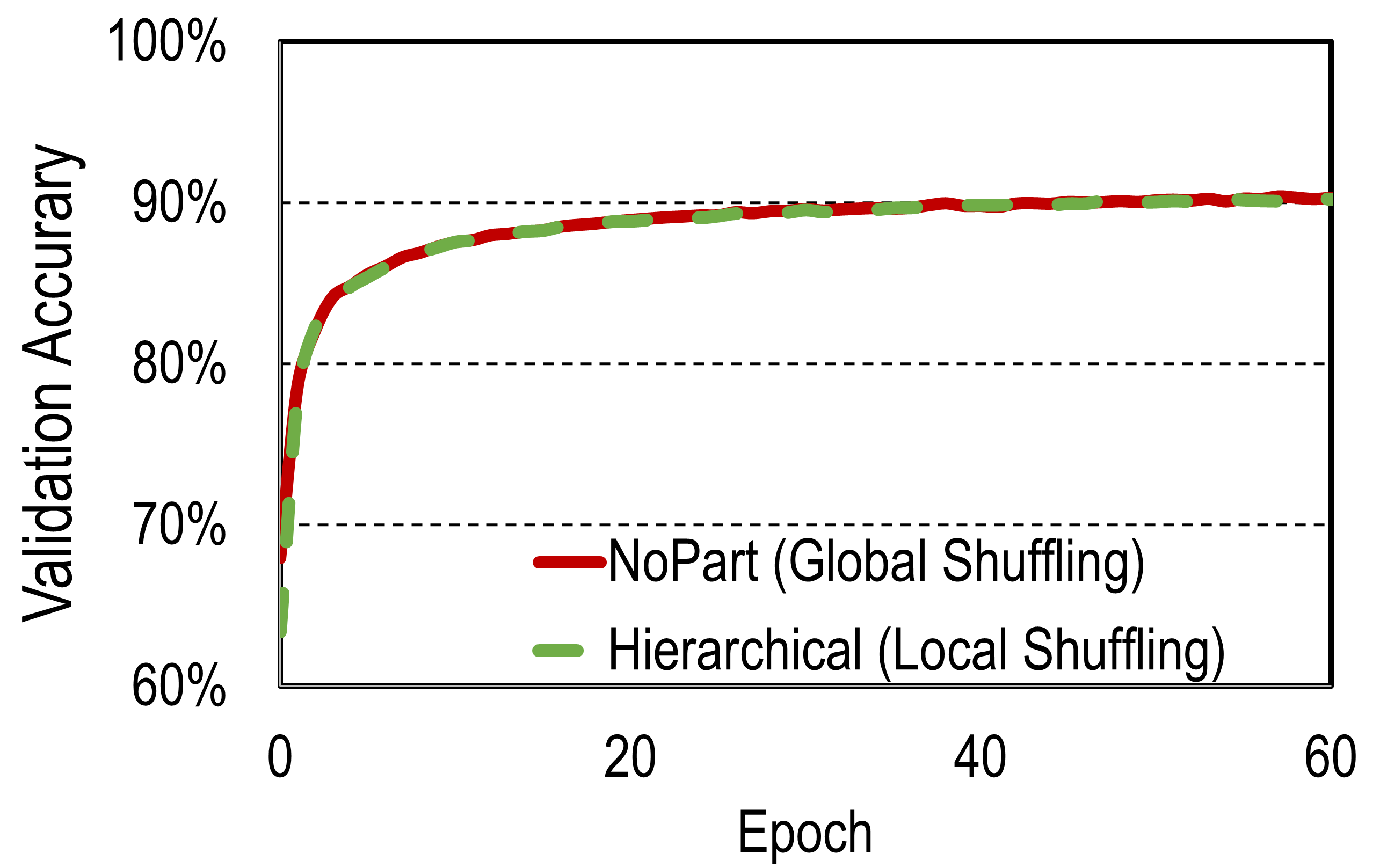}
        \label{convergencePRGCN}
    }
    \vspace{-2ex}
    \caption{Comparing local shuffling and global shuffling on model convergence (NoPart: no partitioning; Hierarchical: hierarchical partitioning).}
    \vspace{-2ex}
    \label{modelconvergence}
\end{figure}

Compared with global shuffling (randomly generating batch seeds from the vertex set of the entire graph),
recent studies~\cite{meng2019convergence, lin2020pagraph} show that local shuffling (generating batch seeds within partitions) brings negligible impact on the rate of model convergence. \SystemName{} adopts local shuffling, and we conduct an experiment on the Siton server (NV2) to compare its convergence speed with global shuffling on both GraphSAGE and GCN using the PR dataset. The results in Figure~\ref{modelconvergence} show that the local shuffling of \SystemName{} could catch up with the convergence speed of global shuffling. 

\subsection{Effect of Unified Cache}
Different from existing cache-based systems, \SystemName{}'s unified cache also takes graph topology into account. In this experiment, we demonstrate the benefits of topology cache. 

We compare the training epoch time of unified cache in \SystemName{} with two baselines: (1) storing all topology in the CPU (denoted as TopoCPU) and (2) replicating the entire topology in every single GPU (denoted as TopoGPU). For a fair comparison, we implement both TopoCPU and TopoGPU in \SystemName{} and use the same GPU memory volume for the three settings. Among the three settings, TopoCPU has the most GPU memory available for the feature cache, and the TopoGPU has the least GPU memory for the feature cache or even runs out of GPU memory. We evaluate PA, CO, and UKS on DGX-V100 and evaluate UKL and CL on DGX-A100. 

\begin{figure}[t]
    \centering
\includegraphics[width=3in]{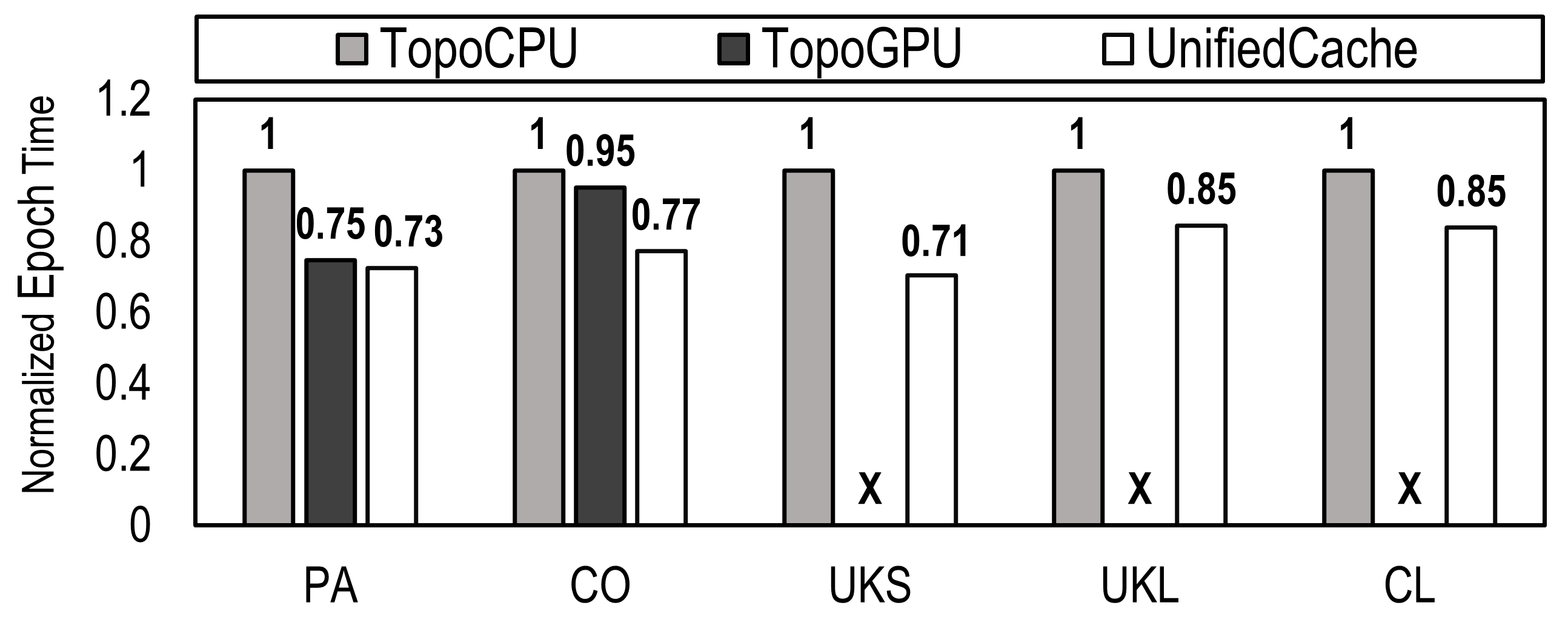} 
	\vspace{-1ex}	
	\caption{The impact of topology cache. ``$\times$'' means OOM (out of memory). } 
	\vspace{-4ex}
	\label{topocache} 
\end{figure}

As shown in Figure~\ref{topocache}, the unified cache outperforms the other two baselines for all graphs. This result demonstrates that, when the size of the feature cache exceeds a threshold, the increase of cache hit rate slows down. In this case, caching some hot topology data in GPU memory will save the system from severe PCIe contention incurred by graph sampling and benefit the overall GNN training throughput.

\vspace{-2ex}
\subsection{Evaluation of Cost Model}
\begin{figure}[t]
\vspace{-1ex}
	\subfloat[PA, Single GPU]
    {\includegraphics[width=1.6in]{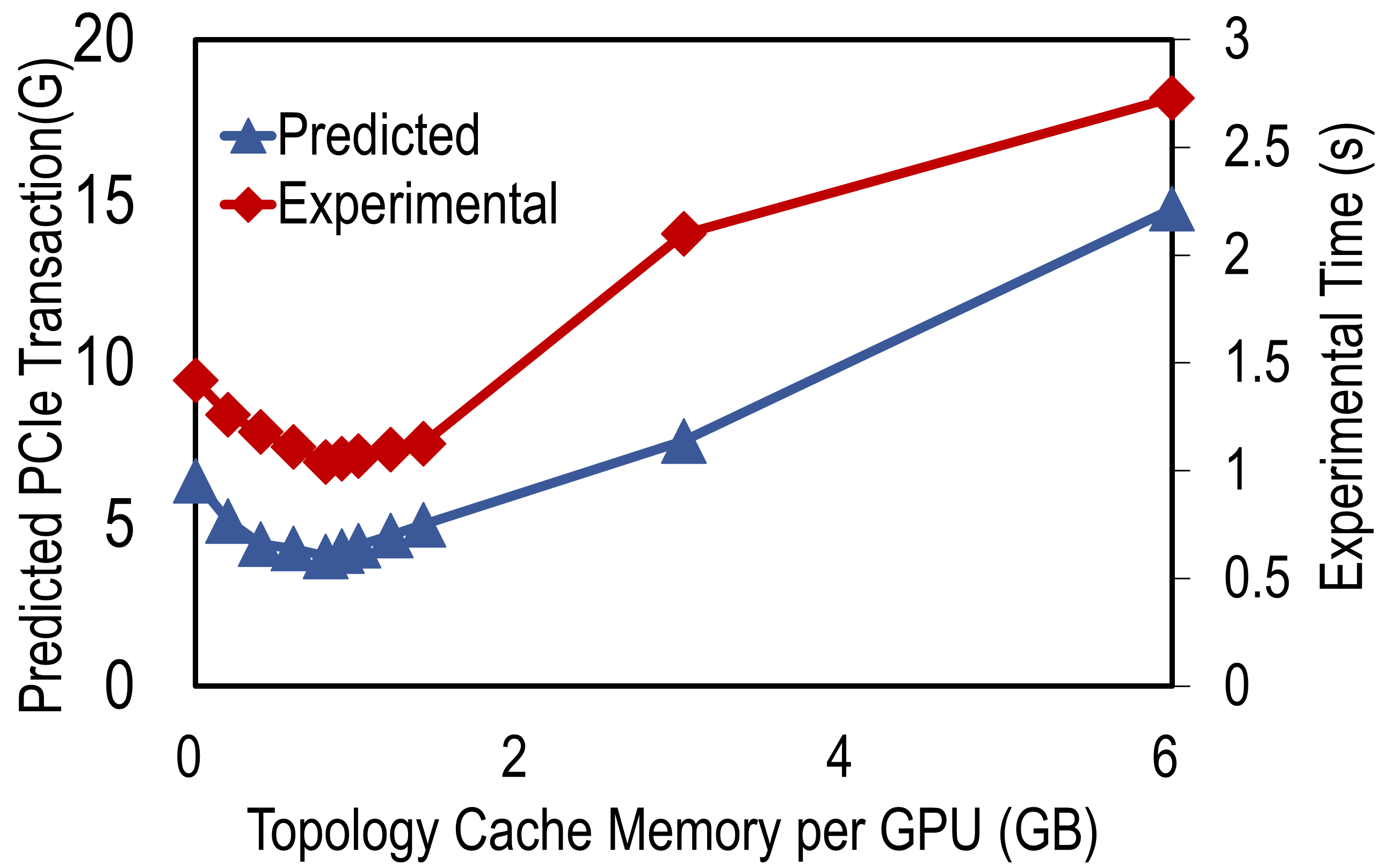} 
    \label{costmodelPA}} 
    \subfloat[UKS, DGX-V100 (NV4)]
    {\includegraphics[width=1.6in]{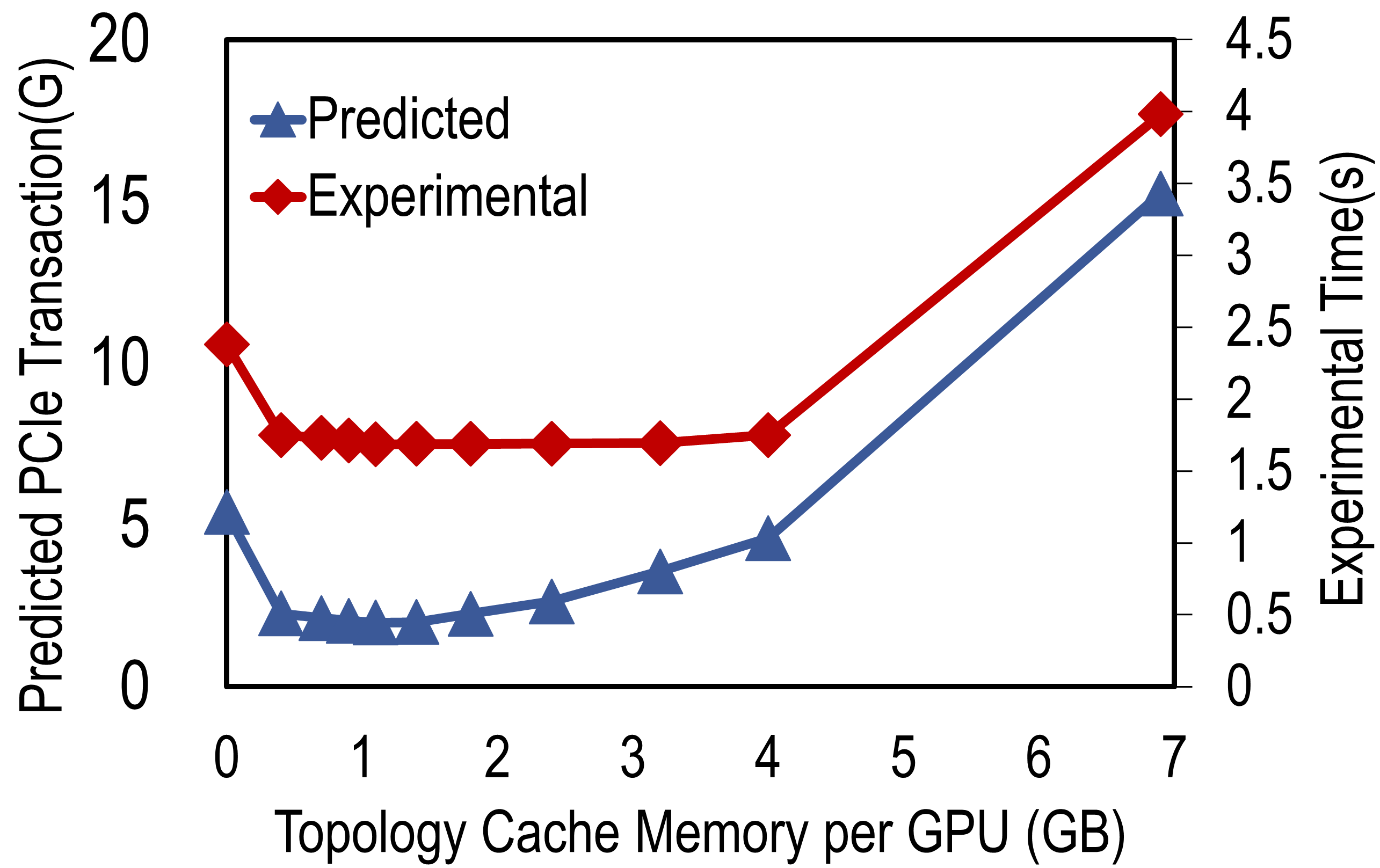} 
    \label{costmodelUKS}}
    \vspace{-2ex}
	\caption{Evaluation of cost model. The left y-axis means the PCIe transaction number predicted by the cost model. The right y-axis represents the experimental per-epoch graph sampling and feature extraction time. }
	\vspace{-2ex}
	\label{costmodel} 
\end{figure}

\SystemName{} proposes the cost model to guide allocating GPU memory for both graph topology and feature cache. In this experiment, we evaluate the effectiveness of this mechanism. Specifically, we compare the predicted PCIe traffic with the experimental per-epoch execution time of graph sampling and feature extraction. In the experiment using the PA dataset, the GPU memory allocated for the cache is $10$ GB. And in the experiment using the UKS dataset, the GPU memory allocated for the cache is $8$ GB. When varying the size of the topology cache, the size of the feature cache is adjusted accordingly. Figure~\ref{costmodel} shows that our cost model can precisely predict the trend of per-epoch execution time without manual interference.

\vspace{-3ex}
\subsection{Partitioning Cost}
\vspace{-1.5ex}
\label{preprocessingcosttxt}
In this experiment, we study the partitioning cost in \SystemName{}. 
We run our experiment on the UKL dataset that has the largest number of edges among all the datasets, resulting in the highest cost of edge-cut partitioning. We also present the results of the PA data (medium size) to show the partitioning costs of different graph scales. We partition PA on DGX-V100 and UKL on Siton using the XtraPulp algorithm. For node classification, we set the training set to be 10\% of the total edges for both graphs. For link prediction, we set the training set to be 80\% of total edges. When the graph is too large to be partitioned in memory, like UKL, we randomly sample a fraction of edges (25\% for UKL) and keep all vertices in the graph such that the subgraph can be partitioned in memory. This technique can obviously speedup graph partitioning 
and preserves a low edge-cut ratio.

Table~\ref{preprocessingcost} shows the preprocessing cost of \SystemName{}'s hierarchical partitioning. We observe that the partitioning cost is tolerable, because 1) we only partition the graph once but can use the partitioning results for multiple GNN training jobs, and 2) the GNN task like link prediction needs multiple epochs to converge while a single epoch often costs a long time to finish. 
\vspace{-2ex}
\begin{table} [t]
\renewcommand\arraystretch{1.4}
\setlength\tabcolsep{3pt}
	\centering

		\begin{scriptsize}
		
	\vspace{-0.5ex}
	\caption{Evaluation of Partitioning Cost.} \vspace{-0.5ex}
	\label{preprocessingcost}
	\begin{tabular}{|c||c|c|}
		\hline
		{\bf Dataset} &  {\bf PA (DGX-V100) }  & {\bf UKL (Siton) } \\
		\hline
            \hline
		{\bf Graph Partition(min)} & 7.2 & 75  \\ 
		\hline
      	{\bf Data Loading From Disk To Memory(min)} & 0.32 & 3.5 \\ 
		\hline
  		{\bf Node Classification Epoch(s)}  & 1.98 & 15.6 \\ 
		\hline
    	{\bf Link Prediction Epoch(min)}  & 49.8 & 402 \\ 
		\hline
	\end{tabular}
	\vspace{-3ex}

		\end{scriptsize}
\end{table}


		


\vspace{5ex}
\vspace{-2ex}
\section{Related Work} 
\vspace{-1.5ex}
To our knowledge, \SystemName{} is the first work that  automatically pushes the envelope of multi-GPU systems for billion-scale GNN training. In the following, we contrast \SystemName{} and existing works in the following aspects. 

\noindent{\textbf{GNN Frameworks. }
Several GNN systems~\cite{jia2020improving, wang2019deep,fey2019fast, zhu2019aligraph, lin2020pagraph, yang2022gnnlab, torchquiver, ma2019neugraph, zhang2020agl, wang2021gnnadvisor, gandhi2021p3, thorpe2021dorylus, zheng2022distributed} have emerged in recent years. 
Most of these GNN systems are built on top of deep learning frameworks like Pytorch~\cite{paszke2019pytorch}, TensorFlow~\cite{abadi2016tensorflow} and MXNet~\cite{chen2015mxnet}. 
}


\noindent{\textbf{GPU Sampling. } NextDoor~\cite{jangda2021accelerating} and C-SAW~\cite{pandey2020c} focus on accelerating GPU sampling kernel. DGL~\cite{wang2019deep} also supports GPU sampling in its recent release. Quiver~\cite{torchquiver} can support GPU sampling with the entire topology either stored in the single GPU or in the CPU memory. GNNLab~\cite{yang2022gnnlab} adopts a factored design where each GPU is dedicated to graph sampling or model training exclusively. In contrast, \SystemName{} uses all GPUs for end-to-end GNN acceleration. 
}

\noindent{\textbf{Graph Partitioning. } 
Graph partitioning such as ~\cite{karypis1997metis, slota2017partitioning, stanton2012streaming, tsourakakis2014fennel, gonzalez2014graphx, gonzalez2012powergraph, petroni2015hdrf, boman2013scalable}, has been widely adopted in GNN systems. DGL~\cite{wang2019deep} adopts METIS~\cite{karypis1997metis} to partition the graph. 
PaGraph~\cite{lin2020pagraph} adopts a self-reliant partitioning strategy with the goal of achieving balanced training vertex allocation across GPUs and improving data locality on every GPU. DGCL~\cite{dgcl_eurosys21} adopts a partitioning algorithm to partition the graph's physical edges and features and store them among distributed machines. In contrast, \SystemName{} adopts hierarchical partitioning to automatically partition graphs to each GPU in a single multi-GPU server accordingly to GPU interconnections.

}

\noindent{\textbf{GPU Feature Cache.} 
PaGraph~\cite{lin2020pagraph}, BGL~\cite{liu2021bgl}, GNNLab~\cite{yang2022gnnlab}, Quiver~\cite{torchquiver} and~\cite{min2022graph} explore feature caching on GPU to accelerate GNN training. PaGraph~\cite{lin2020pagraph} and Quiver~\cite{torchquiver} use the in-degree of vertexes as the hotness metric. 
BGL~\cite{liu2021bgl} applies a FIFO dynamic cache policy and selects training vertices in a BFS order for a higher cache hit rate, but hinders model convergence and incurs cache replacement overheads.
~\cite{min2022graph} uses a weighted reverse PageRank algorithm as a hotness metric. 
GNNLab~\cite{yang2022gnnlab} uses vertices' access frequencies in the pre-sampling epoch as a hotness metric. 
In contrast, \SystemName{} automatically caches both features and topology with the highest hotness. 
And \SystemName{} statically partitions the graph with minimal edge-cut to preserve intra-partition data locality. Figures~\ref{hiclusterexp} and~\ref{modelconvergence} show that Legion can achieve a high cache hit rate even with small cache ratios without compromising the model convergence rate.

\noindent{\textbf{Large Graph Systems.} 
SSD-based GNN systems~\cite{waleffe2023mariusgnn} and distributed GNN systems~\cite{liu2021bgl, zheng2020distdgl, zheng2021distributed, gandhi2021p3} also aim at large-graph training and propose distinct approaches to solve I/O problems at various levels. MariusGNN~\cite{waleffe2023mariusgnn} minimizes I/O between SSD and CPU by including valid graph data in a single swap as much as possible. Systems like BGL~\cite{liu2021bgl}, DistDGLv2~\cite{zheng2021distributed}, and P3~\cite{gandhi2021p3} optimize network I/O between distributed machines, whose network performance can be improved when introducing GPU-centric SmartNIC~\cite{fpganic_atc22}.
In contrast, \SystemName{} focuses on utilizing GPU caches to minimize PCIe traffic from CPU memory to multiple GPUs, which is orthogonal to the above systems.
}

\vspace{-2ex}
\section{Conclusion}
\vspace{-1ex}
We present \SystemName{}, a system that automatically pushes the envelope of multi-GPU systems for billion-scale GNN training. 
 \SystemName{} has three key innovations. First, we propose an NVLink-aware hierarchical partitioning technique that helps minimize cache replication and extends the threshold of cache capacity beyond the limit of a single GPU or NVLink clique. Second, we propose a novel hotness-aware unified cache mechanism that helps accelerate both graph sampling and feature extraction. Third, we present an automatic cache management mechanism enabling optimal cache planning without requiring extra knowledge of hardware specifications and GNN performance details from users. 
Experiments show Legion outperforms SOTA cache-based GNN systems up to 4.32× and supports training on billion-scale graphs.
And \SystemName{} is open-sourced at \url{https://github.com/RC4ML/Legion}.

\noindent {\bf Acknowledgements. }
We thank our shepherd Anand Iyer and anonymous reviewers for their detailed feedback. The work is supported by the following grants: the Program of Zhejiang Province Science and Technology (2022C01044), a research grant from Alibaba Group through the Alibaba Innovative Research (AIR) Program, 
the Fundamental Research Funds for the Central Universities 226-2022-00151, Key Laboratory for Corneal Diseases Research of Zhejiang Province, Starry Night Science
Fund of Zhejiang University Shanghai Institute for Advanced Study (SN-ZJU-SIAS-0010). Zeke Wang and Fei Wu are the corresponding authors.

\bibliographystyle{plain}
\bibliography{references}



\end{document}